%% file: article.tex
\documentclass[preprint,preprintnumbers,a4paper,nofootinbib]{revtex4-1}
\usepackage{amsmath,amssymb}
\usepackage{graphicx}
\usepackage{color}
\usepackage[utf8]{inputenc}
\usepackage{feynmf}

\begin{document}

\begin{flushright}%
{\small EU-TH-103}
\end{flushright}%

\begin{center}
{\Large{\textbf{Numerical analyses of ${\cal N}=2$ supersymmetric quantum mechanics with  cyclic Leibniz rule on lattice}}}
\\
\medskip
\vspace{1cm}
\textbf{
Daisuke Kadoh$^{\,a,b,}$\footnote{kadoh@keio.jp}, 
Takeru Kamei$^{c,}$\footnote{kamei.ehime@gmail.com}, 
Hiroto So$^{d,}$\footnote{hiroto.so@gmail.com}
}
\bigskip

$^a$ {\small Department of Physics, Faculty of Science, Chulalongkorn University, Bangkok 10330, Thailand}

$^b$ {\small Research and Educational Center for Natural Sciences, Keio University, \\ Yokohama 223-8521, Japan}

$^c$ {\small  Graduate School of Science and Engineering, Ehime University, Matsuyama, 790-8577, Japan}

$^d$ {\small  Physics Department, Ehime University, Matsuyama, 790-8577, Japan}

\end{center}

\begin{abstract}
We study a cyclic Leibniz rule, which provides a systematic approach to lattice supersymmetry, 
 using a  numerical method with a transfer matrix. 
 The computation is carried out in ${\cal N}=2$ supersymmetric quantum mechanics 
with the $\phi^6$-interaction for weak and strong couplings.
The computed energy spectra and 
 supersymmetric Ward-Takahashi identities are compared with those obtained 
from another lattice action. We  find that a model with the cyclic Leibniz rule behaves 
similarly to the continuum theory compared with the other lattice action.
\end{abstract}

\maketitle

\input{sec1}

\input{sec2}

\input{sec3}

\input{sec4}

\input{sec5}

\section*{Acknowledgments}
We would like to thank So Matsuura, 
Katsumasa Nakayama and Fumihiko Sugino for their helpful comments.
This work is supported by JSPS KAKENHI Grant Numbers JP16K05328 and 19K03853.

\appendix
\input{sec_app}

\bibliographystyle{unsrt}
\bibliography{refs}

\end{document}

%% file: sec1.tex
\section{Introduction}
\label{introduction}

The difficulty in lattice supersymmetry (SUSY)  is originated from the lack of Leibniz rule \cite{Dondi:1976tx}.
Since any local lattice difference operator does not obey 
Leibniz rule \cite{Kato:2008sp,Kato:2012zh}, 
it is difficult to realize the full SUSY within a local lattice theory \cite{Dondi:1976tx, Kadoh:2009sp, 
Bergner:2009vg, Asaka:2016cxm}. 
Several approaches in which part of SUSY
 is kept on the lattice 
and the full symmetry is restored at the continuum limit have been proposed
so far \cite{ Sakai:1983dg, Catterall:2000rv, Catterall:2001wx, Kikukawa:2002as, 
Cohen:2003xe, Cohen:2003qw, 
Sugino:2003yb, Sugino:2004qd, DAdda:2004dmn, Sugino:2004uv, 
Bergner:2007pu, 
Kadoh:2010ca, Kadoh:2016eju, Schaich:2018mmv}.
Those are, however, the same in a sense that, without getting into details 
about the algebraic structure of a lattice Leibniz rule, 
nilpotent SUSY are realized on the lattice in various ways.
The deep understanding of the lattice Leibniz rule could help us 
to define a lattice model naturally
keeping as many symmetries as possible and 
to study higher dimensional SUSY theories 
without fine tunings, or with less fine tunings.

In Ref.\cite{Kato:2013sba}, another type of the lattice Leibniz rule was proposed 
in ${\cal N}=2$ SUSY quantum mechanics (QM) \cite{Witten:1981nf,Witten:1982df}, which 
keeps a part of symmetries exactly. 
The indices of the new rule appear cyclically
\footnote{
The difference between  the standard Leibniz rule and the cyclic Leibniz rule is shown in section \ref{sec:CLR}.
See  (\ref{LR_2body}) and (\ref{CLR_2body}) for the expressions as a product rule.
} 
and we refer to it as a cyclic Leibniz rule (CLR) 
in this paper as well as the authors of Ref.\cite{Kato:2013sba} did.
The CLR has many solutions and the general solution for a symmetric difference operator
has been studied in Ref.\cite{Kadoh:2015zza}. 
${\cal N}=4$ SUSY QM and  ${\cal N}=2$ SYK model are also defined on the lattice such that 
the half SUSY is exactly kept \cite{Kato:2016fpg, Kato:2018kop}. 
For those models, the exact invariance of half symmetry naturally leads to the CLR 
although there is  another lattice formulation with an exact symmetry
 in ${\cal N}=2$ SUSY QM \cite{Catterall:2000rv}.
Furthermore  a kind of non-renormalization theorem holds for the CLR action of 
the ${\cal N}=4$ case 
such
that any finite correction to the F-term is prohibited \cite{Kato:2016fpg}. 
We can say that  the CLR keeps various natural properties of SUSY at a perturbative level, 
however its non-perturbative property 
which will be important to extend the CLR formulations to higher dimensions is still unknown.

In this paper, we propose a lattice action with the CLR for a backward difference operator
and study its non-perturbative property using numerical computations.
We present a solution of the CLR for any interaction term.
Numerical computations are carried out for the $\phi^6$-interaction for which SUSY is unbroken.
We do not employ the standard Monte-Carlo method 
used in previous studies of SUSY QM
\cite{Catterall:2000rv, 
Giedt:2004vb, Bergner:2007pu, Kanamori:2007ye,  Wozar:2011gu}
but a  direct computational method on the basis of a transfer matrix \cite{Kadoh:2018ele,Kadoh:2018ivg},  
see also \cite{Baumgartner:2014nka, Baumgartner:2015qba, 
Baumgartner:2015zna, Kadoh:2018hqq, Kadoh:2018tis}
for related numerical methods.
The obtained energy spectra show 
that the cut-off dependence of the CLR action is smaller than another lattice action 
defined by Catterall and Gregory (CG) in Ref.\cite{Catterall:2000rv}. 
Numerical results of the SUSY Ward-Takahashi identities (WTIs) also tell us
that full symmetry is restored more rapidly than the CG action 
for the weak and strong couplings.

This paper is organized  as follows. 
In section \ref{sec:susy_quantum_mechanics}, we introduce the continuum and the lattice theories 
of ${\cal N}=2$ SUSY QM.
The continuum theory is given in the Euclidean path integral formulation 
in section \ref{sec:continuum_action} and the lattice theory 
is introduced  in section \ref{sec:lattice_setup}.
The CG lattice action is then presented in section \ref{sec:CG_action}. 
We formulate the CLR for the backward difference operator 
showing a solution for any superpotential and mention a relation 
between the CLR and the standard Leibniz rule in section \ref{sec:CLR_formulation}.
Section \ref{sec:numrical_results} presents the numerical results. 
In section \ref{sec:numrical_method}, we briefly explain the computational method based on
the transfer matrix \cite{Kadoh:2018ele}. 
Then, using computational parameters given in section \ref{sec:parameters},
we show the numerical results of energy spectra in section \ref{sec:energy_spectra}
and those of SUSY WTIs in section \ref{sec:susy_wti}.
We summarize in section \ref{sec:summary}. 
Appendix \ref{more_about_CLR} is devoted to study more about the CLR and 
appendix \ref{sec:weak_coupling_expansion} shows the results of weak coupling expansion
of several lattice actions.

%% file: sec2.tex
\section{SUSY QM and the lattice theory}
\label{sec:susy_quantum_mechanics}

${\cal N}=2$ supersymmetric quantum mechanics is defined in the 
Euclidean path integral formulation according to \cite{Witten:1981nf, Witten:1982df, Cooper:1994eh}. 
We then present a naive lattice approach to SUSY QM and introduce a known improved lattice action 
\cite{Catterall:2000rv}.

\subsection{N=2 SUSY QM}
\label{sec:continuum_action}  
 
With an euclidean time $t$, the action 
of ${\cal N} =2$ SUSY QM is given by 
\begin{equation}
S = \int_0^{\beta}
dt \, \Big\{ \frac{1}{2}({\partial_t}\phi)^2+\frac{1}{2}W^2(\phi) + \bar{\psi} {\partial_t}\psi +  \bar{\psi} W^\prime (\phi) \psi \Big\}, 
\label{N=2action}
\end{equation}
where $\phi(t)$ is a real bosonic variable and $\bar{\psi}(t),\psi(t)$ are one-component
fermionic variables. Those variables satisfy the periodic boundary condition such as $\phi(\beta)=\phi(0)$. 
The superpotential $W(\phi)$ is any function of $\phi$, which determines the physical behavior of this model.
The partition function is defined as
\begin{eqnarray}
Z_P= \int D\phi D\bar \psi D\psi\, e^{-S}
\end{eqnarray}
which is the path integral form of the Witten index.

The classical action is invariant under two SUSY transformations,
\begin{eqnarray}
\begin{split}
&\delta\phi = \epsilon \psi-\bar{\epsilon}\bar{\psi}  \\
&\delta\psi =  \bar{\epsilon}(\partial_t\phi-W) \\
& \delta\bar{\psi} = -\epsilon (\partial_t\phi+W),
\end{split}
\label{N=2}
\end{eqnarray}
where $\epsilon$ and $\bar\epsilon$ are global Grassmann parameters.
The Leibniz rule is needed to show that the action (\ref{N=2action}) is 
invariant under these transformations.

The Witten index $\Delta$ is defined by
\begin{align}
\Delta &\equiv {\rm{Tr}} (e^{-\beta \hat{H}} (-1)^{\hat{F}}), 
\label{Wittenindex}
\end{align}
with the quantum Hamiltonian,  
\begin{align}
\hat H = \frac{1}{2} \hat p^2 + \frac{1}{2}W^2(\hat q) + \frac{1}{2}W'(\hat q) \left[\hat \psi^\dagger, \hat \psi \right],
\label{quantum_Hamiltonian}
\end{align}
where $\hat q$ and $\hat p$ are the position and momentum operator and $\hat \psi^\dag$ and $\hat \psi$
are the creation and annihilation operators, which satisfy $[\hat p,\hat q]=-i$ and $\{\hat \psi,\hat \psi^\dag\}=1$.  
Here $\hat F \equiv \hat \psi^\dag \hat \psi$ 
is the fermion number operator.
The trace  is a summation over all possible normalized states of the system.

We can also write 
\begin{align}
\Delta = {\rm{Tr}} (e^{-\beta \hat{H}_-}) - {\rm{Tr}} (e^{-\beta \hat{H}_+}),
\label{Wittenindex2}
\end{align}
where $\hat{H}_{\pm}=\frac{1}{2} \hat p^2 + \frac{1}{2}W^2(\hat q) \pm \frac{1}{2}W'(\hat q)$ 
are  the Hamiltonians of bosonic $(-)$ and fermionic $(+)$ sectors, respectively.
The Witten index does not depend on $\beta$ because all non-zero eigenmodes in $\hat H_\pm$ form pairs 
and only $\beta$-independent zero modes contribute to $\Delta$.
It is well-known that 
$\Delta$ is zero (non-zero) when SUSY is broken (unbroken) in this model.
We study a SUSY unbroken case with $\Delta= 1$, 
given by $W(\phi) \simeq \lambda \phi^3$ for $|\phi| \rightarrow \infty$
in this paper.

\subsection{Lattice theory}
\label{sec:lattice_setup}
 
The lattice theory is defined on a lattice whose coordinate is  
given by $t = na \, (n \in \mathbb{Z})$.
Lattice bosonic and fermionic variables, which live on the sites, 
are expressed as $\phi_n$ and $\psi_n$, respectively. 
 It is assumed that all variables satisfy the periodic boundary condition,
 \begin{eqnarray}
\phi_{n+N} = \phi_n,\qquad \psi_{n+N} = \psi_n, ,\qquad \bar\psi_{n+N} = \bar\psi_n,
\end{eqnarray}
 where $N$ is the lattice size with $\beta=Na$.

The difference operator $\nabla$ acts on a lattice variable $\varphi_n$ as
$\nabla \varphi_n\equiv \sum_m\nabla_{nm}\varphi_m$
and its transpose is 
$(\nabla^T)_{nm} \equiv \nabla_{mn}$. 
Throughout this paper, $\nabla_+$ and $\nabla_-$ 
denote a simple forward and a backward difference operator, respectively:  
\begin{eqnarray}
&& \nabla_+ \varphi_n \equiv \frac{\varphi_{n+1}-\varphi_n}{a}, 
\label{fwd_op} \\
&& \nabla_- \varphi_n \equiv  \frac{\varphi_{n}-\varphi_{n-1}}{a}.
\label{bwd_op}
\end{eqnarray}
Note that  $(\nabla_+)^T=-\nabla_-$.

The partition function with a lattice action $S$ is defined by
\begin{align}
Z_P & \equiv  \int D\bar{\psi} D\psi D\phi \ e^{-S},
\label{Z_P} 
\end{align}
where
\begin{eqnarray}
&& \int D\phi \equiv  \prod_n \int_{-\infty}^\infty \frac{d\phi_n}{\sqrt{2\pi a}}, 
\label{measure}
\\ 
&& \int D\bar{\psi} D\psi \equiv \int  \prod_n d\bar\psi_n d \psi_n.
\end{eqnarray}
Here each Grassmann measure is an anti-commuting derivative as 
$d\psi_n \equiv \partial/\partial \psi_n$ and  $d\bar\psi_n\equiv  
\partial/\partial \bar{\psi}_n$.

We now consider a naive lattice action,
\begin{equation}
S_{naive}=a \sum_n \Big\{ \frac{1}{2}({\nabla_-}\phi_n)^2+\frac{1}{2} W^2(\phi_n)+
\bar{\psi}_n \nabla_- \psi_n+ \bar\psi_n W'(\phi_n)\psi_n \Big\}
\label{naive_action}
\end{equation}
which is obtained by replacing $\phi(t),\psi(t),\bar\psi(t)$ and $\partial_t$ of (\ref{N=2action})   
by the corresponding lattice variables  $\phi_n,\psi_n,\bar\psi_n$ and $\nabla_-$ 
and replacing the integral by the summation over lattice site.
This action is not invariant under a naive lattice SUSY transformation 
defined by the same replacement of the variables for (\ref{N=2}).

SUSY which is broken at ${\cal O}(a)$ in (\ref{naive_action}) 
is classically restored in the continuum limit $a \rightarrow 0$, 
however such a restoration does not occur at the quantum level.
As seen in later sections,
modifying  ${\cal O}(a)$ interactions of the lattice action, 
we can keep 
only either one of two SUSY transformations 
parametrized by $\epsilon$ and $\bar\epsilon$ at a finite lattice spacing, and 
SUSY is restored in the quantum continuum limit for such a lattice model.

\subsection{Catterall-Gregory lattice model}
\label{sec:CG_action}

Before discussing the CLR,    
we review a lattice  action proposed by Catterall and Gregory \cite{Catterall:2000rv}:
\begin{equation}
S_{CG}=S_{naive} + a \sum_n 
\nabla_- \phi_n W(\phi_n),
\label{S_CG}
\end{equation}
where $\nabla_-$ is the backward difference operator 
defined in (\ref{bwd_op}). 
Note that the added term is a kind of surface 
term which vanishes in the naive continuum limit.

We can show that, in the free limit given by $W(\phi)=ma \phi$,  $S_{CG}$ 
is invariant under the lattice SUSY transformations, 
\begin{eqnarray}
\begin{split}
&\delta\phi_n = \epsilon \psi_n-\bar{\epsilon}\bar{\psi}_n  \\
&\delta\psi_n = \bar{\epsilon}(\nabla_+ \phi_n - W(\phi_n)) \\
&\delta\bar{\psi}_n = -\epsilon (\nabla_- \phi_n+W(\phi_n)). 
\end{split}
\label{lattice_susy_CG}
\end{eqnarray}
For interacting cases, it is not invariant under the whole transformations (\ref{lattice_susy_CG}) 
but invariant under part of SUSY, $\delta_{\epsilon}=\delta|_{\bar\epsilon=0}$: 
\begin{equation}
\delta_{\epsilon}S_{CG}=0.
\end{equation}
This is because the extra term of R.H.S. in (\ref{S_CG}) provides 
$-\delta_\epsilon S_{naive}$ for any finite lattice spacing. 
The remaining $\bar\epsilon$ symmetry in (\ref{lattice_susy_CG}) is restored 
in the quantum continuum limit 
as shown in Refs.\cite{Catterall:2000rv, Giedt:2004vb, Bergner:2007pu, Kadoh:2018ele} 
and also in section \ref{sec:susy_wti} of this paper.

%% file: sec3.tex
\section{Cyclic Leibniz rule for backward difference operator}
\label{sec:CLR_formulation}

We propose an alternative lattice action with the cyclic Leibniz rule (CLR) 
for the backward difference operator and show 
a solution of the CLR for any superpotential. 
It is straightforward to extend the results  
to the case of the forward difference operator.

\subsection{Lattice action with the CLR}
\label{sec:CLR}

The CLR for the symmetric difference operator is proposed in Ref.\cite{Kato:2013sba}.
As an straightforward extension of Ref.\cite{Kato:2013sba}, 
we introduce a lattice action with the CLR for the backward operator:
\begin{equation}
S_{CLR}=a \sum_n \Big\{ \frac{1}{2} ({\nabla_-}\phi_n)^2+\frac{1}{2}( W_n)^2
+\bar{\psi}_n {\nabla_-}\psi_n 
+ \sum_m \bar\psi_n W^\prime_{nm} \psi_m \Big\}, 
\label{CLR_action}
\end{equation}
where $W_n$ is a local function of the boson variables 
\footnote{
Note that $W_n \neq W(\phi_n)$ in general because 
$W_n$ may contain $\phi_m$ with $m \neq n$ as long as the correlation 
rapidly vanishes for $|m-n| \rightarrow \infty$.
See (\ref{locality_M}) of appendix \ref{mbodyCLR} 
for the strict definition of the locality condition.
} 
 and $W^\prime_{nm} \equiv  \frac{\partial W_n}{\partial \phi_m}$.
We now assume that $W_n$ satisfies the CLR,
\begin{equation}
\sum_n \left\{ W_n (\nabla_{-})_{nm}  + \nabla_- \phi_n W^\prime_{nm} \right\}=0. 
\label{CLR2}
\end{equation}
As explained in the next section, 
a desirable local solution is 
\begin{eqnarray}
W_{n}
= \frac{U(\phi_{n})-U(\phi_{n - 1})}{\phi_{n}-\phi_{n - 1}}, 
\label{generic-potential}
\end{eqnarray}
where $U (\phi) = \int^{\phi} d\phi^\prime  \, W(\phi^\prime)$. 
The lattice action (\ref{CLR_action}) classically reproduces the continuum one
(\ref{N=2action}) as $a \rightarrow 0$ since $W_n=W(\phi_n) + {\cal O}(a)$.

The importance of CLR is understood 
by considering a half lattice SUSY transformation,
\begin{eqnarray}
\begin{split}
&\delta_\epsilon \phi_n = \epsilon \psi_n \\
&\delta_\epsilon\psi_n = 0 \\
&\delta_\epsilon \bar{\psi}_n = -\epsilon (\nabla_-\phi_n + W_n). 
\end{split}
\label{lattice_susy_CLR}
\end{eqnarray}
The lattice action  (\ref{CLR_action}) with any solution of (\ref{CLR2})
is invariant under (\ref{lattice_susy_CLR})
because 
\begin{eqnarray}
\delta_\epsilon S_{CLR} =  \epsilon  a \sum_{n} X_n \psi_n=0,
\end{eqnarray}
where 
\begin{eqnarray}
X_n \equiv  -\sum_m \left\{ W_m (\nabla_-)_{mn} + W^\prime_{mn} \nabla_-\phi_m \right\} 
\end{eqnarray}
which vanishes as long as $ W_n$ satisfies the CLR (\ref{CLR2}).

The other half transformation of 
 ${\cal N}=2$ is broken on the lattice in general, which 
is restored at the continuum limit 
as seen in section \ref{sec:susy_wti}. 
However, in the free theory, 
it still remains as an exact symmetry because 
the free lattice action with the solution (\ref{generic-potential}) 
is invariant under
\begin{eqnarray}
\begin{split}
& \delta_{\bar{\epsilon}}\phi_n = -\bar{\epsilon} \bar{\psi}_n \\
& \delta_{\bar{\epsilon}}\psi_n = \bar{\epsilon} ( \nabla_+ \phi_n -W_{n+1} )
\\
& \delta_{\bar{\epsilon}}\bar{\psi}_n = 0.
\end{split}
\label{lattice_CLR_another}
\end{eqnarray}
Note that $W_{n+1}$ is used in  $\delta_{\bar{\epsilon}}\psi_n$ instead of  $W_{n}$.
We can actually show that 
\begin{eqnarray}
\delta_{\bar\epsilon} S_{CLR} =  \bar\epsilon \left\{
a\sum_{n}  X_n \bar \psi_n 
+ a\sum_{n,m}  Y_{nm} (W_n \bar \psi_m - \bar\psi_m \nabla_-\phi_n)
+ a\sum_{nmk} Z_{nmk}  \bar \psi_n  \bar \psi_k  \psi_m
 \right\},
 \label{dS_CLR}
\end{eqnarray}
where 
\begin{eqnarray}
\begin{split}
& Y_{nm} \equiv W^\prime_{m,n-1} -W^\prime_{n,m} \\
& Z_{nmk} \equiv \frac{\partial^2 W_n}{\partial \phi_k \partial \phi_m}.
\end{split}
\end{eqnarray}
Although we have $X_n=0$ from the CLR, $Y_{nm}$ and $Z_{nmk}$ do not vanish 
for a generic superpotential.
However, for the free theory with the solution (\ref{generic-potential}),  
\begin{eqnarray}
W_n = \frac{m}{2} (\phi_n + \phi_{n-1}),
\label{1body_W}
\end{eqnarray}
it is easy to show that $Y_{mn}$, $Z_{nmk}$ and (\ref{dS_CLR}) vanish.

\subsection{A solution of CLR for the backward difference operators}
\label{sec:CLR_solution}  

We show that  (\ref{generic-potential}) is a local and well-defined solution of (\ref{CLR2}) 
for a generic superpotential. 
Once the solution is given, the lattice CLR action 
retains an exact SUSY as seen in the previous section.

Let us first consider the free theory.
For the backward operator $(a\nabla_-)_{nm}=\delta_{nm} -\delta_{n-1,m}$,
we take an ansatz solution within the nearest neighbor interactions, 
 $W_n=d_0 \phi_n + d_1\phi_{n-1} + d_2 \phi_{n+1}$.   
It is then found that $d_0=d_1=1/2, d_2=0$ is  a solution of (\ref{CLR2}), for which  
(\ref{1body_W}) is obtained.

It is not easy to apply such a straightforward 
way to a generic superpotential. 
We derive another representation of (\ref{CLR2}) to find a solution.
Rescaling $\phi_n$ of  (\ref{CLR2}) as $u \phi_n$ 
with a parameter $u \in [0,1]$ and using the chain rule for $\partial_u$,
we obtain 
\begin{eqnarray}
\frac{\partial}{\partial u} \sum_n \left\{  u \nabla_- \phi_n 
W_n|_{\phi \rightarrow u \phi} \right\} =0.
\label{dif_CLR}
\end{eqnarray}
Integrating (\ref{dif_CLR}) from $u=0$ to $u=1$, 
we find a condition that means a vanishing surface term, 
\begin{equation}
\sum_n \nabla_- \phi_n W_n =0. 
\label{CLR3}
\end{equation}
This condition is equivalent to (\ref{CLR2}) 
because  (\ref{CLR2}) can also be derived  from (\ref{CLR3}) 
differentiating  (\ref{CLR3}) with respect to $\phi_m$.

The relation
(\ref{CLR3}) is easily solved  by a local function (\ref{generic-potential}). 
All we have to do is check whether or not $W_n$ given by  (\ref{generic-potential}) is a well-defined function
that coincides with $W(\phi_n)$ as $a\rightarrow 0$. 
By integrating  
$\partial_u U(\phi_{n} - u a\nabla_- \phi_n)$ from $u=0$ to $u=1$ and using the chain rule for $\partial_u$, 
we have 
\begin{eqnarray}
U(\phi_{n}) - U(\phi_{n-1}) =  (\phi_n-\phi_{n -1}) \int_0^1 du\,  W(\phi_{n} - au \nabla_- \phi_n).
\end{eqnarray}
The division in (\ref{generic-potential}) is well-defined 
because the integral of R.H.S. is well-defined for any configuration of $\phi_m$. 
Since  the integral is $W(\phi_n)$ up to ${\cal O}(a)$, 
we can immediately show that $ W_{n} = W(\phi_n) + {\cal O}(a)$.

\subsection{CLR v.s. Leibniz rule}
\label{sec:relation_to_LR}

The difference between the CLR and the standard Leibniz rule (LR) is discussed here. 
In the continuum theory, 
LR for $\partial_t$ is $\partial_t W(\phi) = W^\prime (\phi) \partial_t \phi$.
So 
a naive lattice LR  is introduced as
\begin{eqnarray}
{\rm LR}: \quad   \sum_m \{ \nabla_{nm} W_m - W^\prime_{nm} (\nabla \phi)_m\}=0,
\label{general_LR}
\end{eqnarray}  
for $W_n$ that is a local function of bosonic variables.
Here we again use $W^\prime_{nm} \equiv \partial W_n/\partial \phi_m$.
We find that the CLR is different from LR in general since
\begin{eqnarray}
{\rm CLR}: \quad  \sum_m \{ -\nabla^T_{nm} W_m - W^\prime_{mn} (\nabla \phi)_m \}=0.
\label{general_CLR}
\end{eqnarray}  
Note that $W^\prime$ in the second term is transposed.

The CLR coincides with LR if $W^\prime_{nm}=W^\prime_{mn}$ 
for $\nabla^T=-\nabla$ (symmetric difference operators), which corresponds to
the case that the lattice action is invariant under both of two SUSY transformations \cite{Kato:2013sba}.
However, the no-go theorem \cite{Kato:2008sp} tells us that
LR does not hold for any difference operator and any interacting cases with keeping the locality principle. 
It is therefore difficult to realize the full SUSY 
 transformation exactly on the lattice. 
The CLR cannot be realized  with a non-trivial solution in this case.

The similar argument holds for the backward difference operator $\nabla_-$. 
Suppose that $W_n$ is a solution of the CLR and $\delta_{\bar\epsilon} S=0$.
Using $W^\prime_{mn}=W^\prime_{n,m-1}$ from $Y_{nm}=0$,
we can show that the CLR coincides with LR for $\nabla_+$ 
since $\nabla_-^T=-\nabla_+$ and 
$\sum_m W^\prime_{mn} (\nabla_- \phi)_{m} = \sum W^\prime_{nm} (\nabla_+ \phi)_m$.
The no-go theorem again tells us that one cannot find a solution of the CLR 
so that the lattice action (\ref{CLR_action}) is invariant under both of $\delta_\epsilon$ and $\delta_{\bar\epsilon}$.

The lattice rules (\ref{general_LR})  and (\ref{general_CLR}) can 
also be expressed as  a product rule of lattice variables. 
As an example, let us consider a lattice superpotential, 
\begin{eqnarray}
W^{e.g.}_n \equiv \sum_{m,k} M_{nmk} \phi_m \phi_k, 
\end{eqnarray} 
as a discretization of $W^{e.g.}(\phi(x)) = \phi^2(x)$. 
Then the (two-body) LR can be expressed as
\begin{eqnarray}
\sum_{n}\Big\{ \nabla_{na} M_{bnc}  - \nabla_{bn} M_{nca} + \nabla_{nc} M_{ban} \Big\}=0,
\label{LR_2body}
\end{eqnarray}
while the (two-body) CLR is 
\begin{eqnarray}
\sum_{n}\Big\{\nabla_{na} M_{nbc}+\nabla_{n b} M_{nca} + \nabla_{nc} M_{nab} \Big\} =0 .
\label{CLR_2body}
\end{eqnarray}
The name of {\it cyclic} Leibniz rule comes from a cyclicity of the indices $a,b,c$.    
In appendix \ref{more_about_CLR}, an explicit solution for the $m$-body CLR is also given.

%% file: sec4.tex
\section{Numerical results}
\label{sec:numrical_results}

Numerical computation
is carried out for the CLR action (\ref{CLR_action}) with the periodic boundary conditions
for the superpotential,
\begin{equation}
W=m\phi+\lambda m^2 \phi^3,  
\label{phi3_potential}
\end{equation}
where $\lambda$ is the dimensionless coupling constant and $m$ is the mass.
Supersymmetry is kept unbroken since  the Witten index is nonzero for this potential.   
The energy spectra and the SUSY Ward Takahashi identities are evaluated 
at two coupling constants $\lambda=0.001$ (weak) and $\lambda=1$ (strong).
We compare the results with those obtained from the CG action (\ref{S_CG}) to understand 
the dependence of the results on the lattice spacing.

\subsection{Numerical methods}
\label{sec:numrical_method}

We begin with giving the CLR lattice action used in the actual computations: 
\begin{equation}
S_{CLR}=a \sum_n \Big\{ \frac{1}{2} ({\nabla_-}\phi_n)^2+\frac{1}{2}
(W_n)^2+\bar{\psi}_n {\nabla_-}\psi_n 
+ \sum_m \bar\psi_n W^\prime_{nm} \psi_m \Big\}, 
\label{N=2CLRaction}
\end{equation}
\noindent
where
\begin{eqnarray}
&& W_n=\frac{ma}{2}( \phi_n + \phi_{n-1}) + \frac{(ma)^2 \lambda}{4} 
 (\phi_n^3 + \phi_n^2 \phi_{n-1} + \phi_n \phi^2_{n-1} + \phi^{3}_{n-1}),
\label{superpotW}
\\
&&
W^\prime_{nm} =  \frac{ma}{2}( \delta_{nm} +\delta_{n-1,m}) 
+\frac{(ma)^2 \lambda}{4} \big\{ (3\phi_n^2+2\phi_n \phi_{n-1}+\phi^2_{n-1}) \delta_{nm} 
\nonumber \\
 && \hspace{2cm}
 +(\phi_n^2+2\phi_n \phi_{n-1}+3\phi^2_{n-1})\delta_{n-1,m} \big\}.
\label{superpotW'}
\end{eqnarray}
As shown in section \ref{sec:CLR},
the action (\ref{N=2CLRaction}) is invariant under a single SUSY transformation (\ref{lattice_susy_CLR})
thanks to the CLR (\ref{CLR2}).

The partition function and the correlation functions are expressed in terms of 
transfer matrices. It it straightforward to show that,
integrating out the fermionic variables, the partition function (\ref{Z_P}) with (\ref{N=2CLRaction}) 
is given as 
\begin{equation}
Z_P=\int D\phi \, \left\{ \prod_{n=1}^N (1+A_{\phi_n \phi_{n-1}})e^{-{\cal L}_{\phi_n\phi_{n-1}}} 
- \prod_{n=1}^N (1-A_{\phi_{n-1}\phi_n})e^{-{\cal L}_{\phi_n\phi_{n-1}}}  \right\},
\label{Z_CLR_intF}
\end{equation}
where 
\begin{eqnarray}
&& A_{\alpha\beta} \equiv  \frac{ma}{2} +\frac{(ma)^2\lambda}{4}(3\alpha^2+2\alpha\beta + \beta^2),\\
&& {\cal L}_{\alpha\beta} \equiv \frac{1}{2} (\alpha-\beta)^2 + \frac{1}{8} \left( 
ma( \alpha + \beta) + \frac{(ma)^2 \lambda}{2} 
 (\alpha^3 + \alpha^2 \beta + \alpha \beta^2 + \beta^3 )
\right)^2, 
\end{eqnarray}
because $S_B = \sum_{n=1}^N {\cal L}_{\phi_n\phi_{n-1}}$ and 
$W^\prime_{nm} = A_{\phi_n\phi_{n-1}} \delta_{n,m}  + A_{\phi_{n-1}\phi_n}\delta_{n-1,m} $. 
Note that $A_{\alpha\beta}$ and ${\cal L}_{\alpha\beta}$ are  infinite dimensional matrices 
since $\alpha,\beta \in {\mathbb R}$.

In order to define finite dimensional matrices, 
each path integral measure of ($\ref{Z_CLR_intF}$) is discretized 
by the Gauss-Hermite quadrature.  For a function $f(x)$, 
the Gauss-Hermite quadrature formula is given by an approximation of the integral:
\begin{equation}
\int_{-\infty}^{\infty} dx f(x) \approx \sum_{x \in S_K}  g_K(x) f(x),
\label{GH_method}
\end{equation}
where $S_K$ is a set of roots of $K$-th Hermite polynomial $H_K$ and the weight $g_K(x)$ is
\begin{equation}
g_K(x) = \frac{2^{K-1} K! \sqrt{\pi}}{K^2 H^2_{K-1}(x)} e^{x^2}.
\end{equation}
Since $K$ is the order of the approximation, 
the sum of (\ref{GH_method}) is expected to reproduce the integral of L.H.S. 
as $K \rightarrow \infty$.

We can express $Z_P$ using finite dimensional matrices $T_\pm$ as  
\begin{align}
Z_P \approx  {\rm{tr}} (T_-^N) - {\rm{tr}} (T_+^N),
\label{finite_Z}
\end{align}
discretizing all path integral measures (\ref{measure}) by the quadrature: 
\begin{equation}
\int D\phi \approx 
\frac{1}{(2\pi)^{N/2}} 
\sum_{\phi_1 \in S_K} 
\cdots \sum_{\phi_N \in S_K} \ g_K(\phi_1) \cdots g_K(\phi_N).
\end{equation}
Here, for $\alpha,\beta \in S_K$,
\begin{align}
& (T_{-})_{\alpha\beta} \equiv (1 + A_{\alpha\beta} ) R_{\alpha\beta},  \label{T_+} \\
&  (T_{+})_{\alpha\beta} \equiv  (1 - A_{\beta\alpha}) R_{\alpha\beta},  \label{T_-}  \\
& R_{\alpha\beta} \equiv \sqrt{\frac{g_K(\alpha) g_K(\beta)}{2\pi}} \ e^{ -{\cal L}_{\alpha\beta}}.
\end{align}
A comparison with (\ref{Wittenindex2}) tells us that
$T_-$ and $T_+$ are a bosonic and fermionic 
transfer matrix, respectively. 
The trace of (\ref{finite_Z}) means 
\begin{align}
{\rm tr}(X) \equiv  \sum_{\alpha \in S_K} X_{\alpha\alpha},
\label{def_trace}
\end{align}
where $X_{\alpha\beta}$ is a matrix with $\alpha,\beta \in S_K$.

Similarly, any correlation function is given in terms of the transfer matrices. 
We basically follow Ref.\cite{Kadoh:2018ele} to derive the expressions.
The two point correlation function of the bosonic variable is 
\begin{equation}
\langle  \phi_j \phi_k  \rangle \approx \frac{1}{Z} \ 
 {\rm Tr} \bigg\{ T_-^{N-k+j} D T_-^{k-j} D- T_+^{N-j+k}D T_+^{k-j} D \bigg\},
 \label{boson-correlator}
\end{equation}
for $0 \le j \le k \le N$. Here $D$ represents an operator insertion, which is defined as 
\begin{equation}
D_{\alpha\beta} \equiv \alpha \delta_{\alpha\beta}.
\end{equation}
The boson two-point function is exactly the same formula as 
that of the CG action \cite{Kadoh:2018ele}.  
On the other hand,  the fermion two-point function is 
slightly different:
\begin{equation}
\langle \psi_{j} \bar{\psi}_{k}\rangle
\approx \frac{1}{Z} \
{\rm tr~}  \bigg\{ R T_-^{k-j-1}T_+^{N+j-k}  \bigg\}. 
\label{fermi-correlator}
\end{equation}
for $0 \le j \le k \le N$. 
\footnote{The formula for the CG action given in  \cite{Kadoh:2018ele} is reproduced 
because $T_+ = R$ in the case. }

The transfer matrices $T_\pm$ can be improved by rescaling the bosonic variables before the discretization of the measures.
According to Ref.\cite{Kadoh:2018ele}, applying the quadrature after
rescaling $\phi$ as $\phi \to \phi/s \ (s \in \mathbb{R})$,
we have
\begin{align}
& (T^{(s)}_{-})_{\alpha\beta} \equiv (1 + A^{(s)}_{\alpha\beta} ) R^{(s)}_{\alpha\beta},  \\
&  (T^{(s)}_{+})_{\alpha\beta} \equiv  (1 - A^{(s)}_{\beta\alpha}) R^{(s)}_{\alpha\beta},  \\
& R^{(s)}_{\alpha\beta} \equiv \sqrt{\frac{g_K(\alpha) g_K(\beta)}{2\pi s^2}} \ e^{ -{\cal L}^{(s)}_{\alpha\beta}}.
\end{align}
where $A^{(s)}_{\alpha\beta} \equiv A_{\alpha(s)\,\beta(s)}$ 
and ${\cal L}^{(s)}_{\alpha\beta} \equiv {\cal L}_{\alpha(s)\,\beta(s)}$ with $\alpha(s) \equiv \alpha/s$ and $\beta(s) \equiv  \beta/s$. 
The partition function and the correlation functions are then given 
by the same formulas as (\ref{finite_Z}), (\ref{boson-correlator})
and  (\ref{fermi-correlator})
with $T_\pm^{(s)}$ 
and $R^{(s)}$ instead of $T_\pm$ and $R$.
The operator insertion $D$ is also replaced by $D^{(s)} =D/s$. 
The trace is still given by (\ref{def_trace}). 
We can obtain computational results with a high precision by tuning the rescaling parameter $s$
 such that the Witten index $Z_P=1$ is realized as accurate as possible.

\subsection{Computational parameters}
\label{sec:parameters}

Table \ref{tab:parameters} shows the parameters used in our computations 
of the CLR action. 
We employ two representative coupling constants,  
$\lambda=0.001$ as a weak coupling and
$\lambda=1$ as a strong coupling.  
The rescaling parameter $s$ should be tuned
for each parameter set such that  the Witten index $Z_P=1$ is reproduced 
as accurate as possible, as done in Ref.\cite{Kadoh:2018ele}.  
The matrix sizes $K$ used for the SUSY WTI are smaller than those for
 the mass spectra to reduce the computational cost.   
This is because the SUSY WTIs are evaluated by performing the direct matrix product several times 
while the mass spectra are evaluated by diagonalizing $T_\pm$ once. 
Similarly, we use the same lattice sizes with a slightly different $s$ for the CG action.

We take $m\beta=30$ that is large enough to obtain the numerical results 
with a negligible finite $\beta$ effect because 
$e^{-\beta E_1} < O(10^{-13})$ for the first excited energy $E_1/m \ge  1$. 
The lattice spacing
is  shown as rounded numbers, which is 
uniquely determined from the lattice size $N$ for fixed $m\beta$ as $ma=m \beta/N (=30/N)$.
For instance,  $ma=0.017964\ldots$  for $N=1670$ is denoted as $ma=0.018$ in the table 
but we use $ma=30/N$ in the actual computations without loss of digit.

\begin{figure}[!htbp]
\includegraphics[]{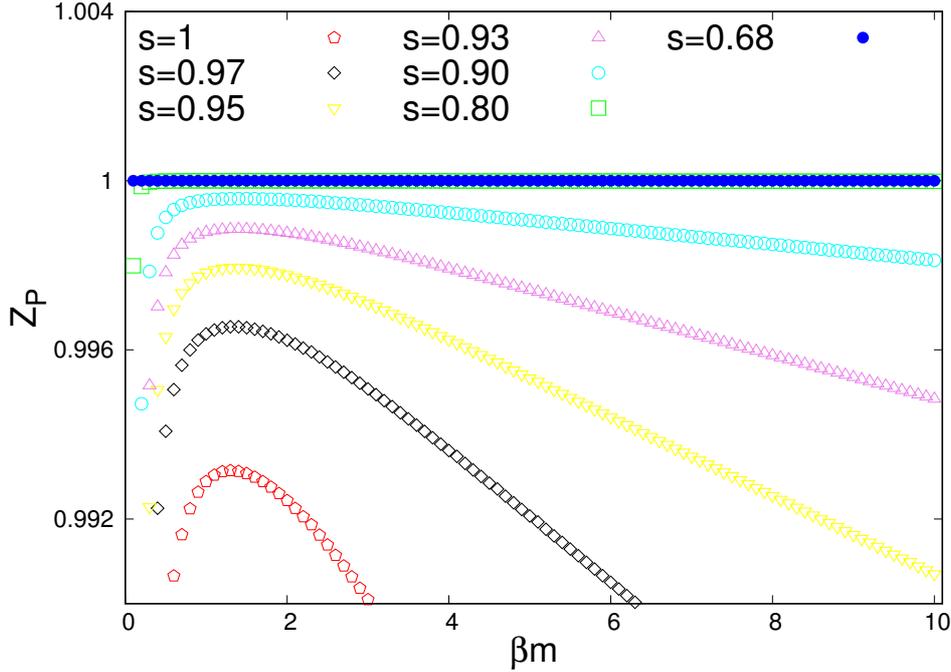}
\caption{Partition function with the periodic boundary condition against $\beta$ for the CLR action.
             We use several $s$ with fixed $K=150$ for $ma=0.01, \lambda=1$. }
\label{fig:Witten_index}
\end{figure}

Figure \ref{fig:Witten_index} shows the results of $Z_P$ against $\beta$ for several $s$. 
Although $Z_P$ is analytically shown to be unity even on the lattice \cite{Kato:2013sba}, 
the numerical results depend on $\beta$.
The deviations from $Z_P=1$ are systematic errors
which come from the finite $K$-effect.
We can decrease the errors tuning $s$ for fixed $K$.
We find that $s=0.68$ leads to $|Z_P - 1| < O(10^{-9})$ for $K=150$ 
in the case of $ma=0.01$ and $\lambda=1$.
Each parameter has 
a different value of $s$ so that $Z_P=1$ is realized within $O(10^{-9})$ 
as shown in Table \ref{tab:parameters}.

\begin{table}[!htbp]
\begin{center}
\begin{tabular}{c}
\begin{minipage}{1\hsize}
\begin{center}
\begin{tabular}{c c c c | c c c c} \hline \hline
\multicolumn{8}{c}{$\lambda=0.001$} \\ \hline \hline
\multicolumn{4}{c}{Energy spectra} & \multicolumn{4}{c}{SUSY WTI} \\ \hline \hline
am & s & N & K & am & s & N & K \\ \hline
0.020 & 0.47 & 1500 & \ 150 \ & \ 0.600 \ & 1.39 & 50 & 40\\
0.019 & 0.46 & 1580 & 150 & \ 0.500 \ & 1.26 & 60 & 40   \\
0.018 & 0.45 & 1670 & 150 & 0.400 & 1.13 & 75 & 40 \\
0.017 & 0.44 & 1770 & 150 & 0.300 & 0.97 & 100 & 40 \\
0.016 & 0.42 & 1880 & 150 & 0.250 & 0.89 & 120 & 40 \\
0.015 & 0.41 & 2000 & 150 & 0.200 & 0.79 & 150 & 40 \\
0.014 & 0.40 & 2140 & 150 & 0.150 & 0.68 & 200 & 40 \\
0.013 & 0.38 & 2310 & 150 & 0.100 & 0.56 & 300 & 40 \\
0.012 & 0.37 & 2500 & 150 & 0.080 & 0.51 & 375 & 40 \\
0.011 & 0.36 & 2730 & 150 & 0.060 & 0.49 & 500 & 50 \\
0.010 & 0.34 & 3000 & 150 & 0.050 & 0.44 & 600 &  50 \\
0.009 & 0.33 & 3330 & 150 & 0.040 & 0.44 & 750 &  60 \\
0.008 & 0.33 & 3750 & 170 & 0.030 & 0.41 & 1000 & 70 \\
0.007 & 0.30 & 4290 & 170 & 0.025 & 0.34 & 1200 & 70 \\
0.006 & 0.27 & 5000 & 170 & 0.020 & 0.32 & 1500 & 80 \\
0.005 & 0.27 & 6000 & 200 & 0.015 & 0.33 & 2000 & 100 \\
0.004 & 0.24 & 7500 & 200 & 0.010 & 0.29 & 3000 & 120 \\ \hline \hline
\multicolumn{1}{c}{} \\
\multicolumn{1}{c}{} \\
\multicolumn{1}{c}{} \\ 
\end{tabular}
\hspace{0.5cm}
\begin{tabular}{c c c c | c c c c} \hline \hline
\multicolumn{8}{c}{$\lambda=1$} \\ \hline \hline
\multicolumn{4}{c}{Energy spectra} & \multicolumn{4}{c}{SUSY WTI} \\ \hline \hline
am & s & N & K & am & s & N & K \\ \hline
0.020 & 0.97 & 1500 & \ 150 \ & \ 0.600 \ & 2.93 & 50 & 40\\
0.019 & 0.95 & 1580 &  150  & 0.500 & 2.68 & 60 & 40   \\
0.018 & 0.92 & 1670 &  150   & 0.400 & 2.46 & 75 & 40 \\
0.017 & 0.90 & 1770 &  150   & 0.300 & 2.08 & 100 & 40 \\
0.016 & 0.87 & 1880&    150  & 0.250 & 1.89 & 120 & 40 \\
0.015 & 0.84 & 2000 &   150  & 0.200 & 1.69 & 150 & 40 \\
0.014 & 0.81 & 2140 &   150   & 0.150 & 1.47 & 200 & 40 \\
0.013 & 0.78 & 2310 &   150   & 0.100 & 1.18 & 300 & 40 \\
0.012 & 0.75 & 2500 &  150    & 0.080 & 1.06 & 375 & 40 \\
0.011 & 0.72 & 2730 &  150    & 0.060 & 0.91 & 500 & 40 \\
0.010 & 0.68 & 3000 &  150   & 0.050 & 0.83 & 600 &  40 \\
0.009 & 0.65 & 3330 &   150  & 0.040 & 0.74 & 750 &  40 \\
0.008 & 0.61 & 3750 & 150    & 0.030 & 0.64 & 1000 & 40 \\
0.007 & 0.57 & 4290 &   150  & 0.025 & 0.65 & 1200 & 50 \\
0.006 & 0.53 & 5000 &    150 & 0.020 & 0.57 & 1500 & 50 \\
0.005 & 0.48 & 6000 &  150   & 0.015 & 0.58 & 2000 & 70 \\
0.004 & 0.46 & 7500 &  170   & 0.010 & 0.47 & 3000 & 70 \\
0.003 & 0.40 & 10000 &  170 &&&&   \\
0.002 & 0.32 & 15000 &  170 &&&&  \\
0.001 & 0.23 & 30000 &  200 &&&&\\  \hline \hline
\end{tabular}
\caption{Parameters used in the numerical computations of the CLR system.
             Left and right tables  are ones for a weak coupling $\lambda=0.001$ and for
             a strong coupling $\lambda=1$, respectively.}
\label{tab:parameters}
\end{center}
\end{minipage}
\end{tabular}
\end{center}
\end{table}

\subsection{Energy spectra}
\label{sec:energy_spectra}

The energy spectra of lattice SUSY quantum mechanics are read from two transfer 
matrices $T_\pm$ associated with two Hamiltonians $\hat H_\pm $ as 
$T_\pm \approx e^{-a \hat H_\pm}$.
The energy eigenvalues of the bosonic and fermionic states 
$E_n^B$ and $E_n^F$ are thus obtained from the $n$-th eigenvalue of $T_\pm$:  
$(T_-)_n= {\rm e}^{- a E_n^B }$ and $(T_+)_n = {\rm e}^{-a E_n^F}$. We use  
 numerical diagonalizations of $T_\pm$ to evaluate $(T_\pm)_n$.
The non-zero eigenvalues are degenerate between $\hat H_+$ and $\hat H_-$ 
and only  $\hat H_-$ has a zero mode for the superpotential (\ref{phi3_potential}).
We expect that $T_\pm$ have the same spectra
even on the lattice thanks to the exact SUSY.

\subsubsection{Weak coupling results}

Table \ref{tab:energy_spectra_weak} shows the
ten smallest energy eigenvalues obtained from the CLR action 
for $\lambda=0.001$ at a lattice spacing $ma=0.01$.
The central values are ones obtained for $K=150$ and 
the errors are estimated from the largest difference among the results with $K=140,150,\ldots,200$.
The spectra look like ones of the harmonic oscillator, $E_n=n m\, (n=1,2,\cdots)$, 
since $\lambda=0.001$ is sufficiently small. 
As we expected, $E^{B}_n$ and $E^{F}_n$ coincide with each other within the errors. 
The same degeneracies are observed for the other lattice spacings.

\begin{table}[!htbp]
\begin{center}
\begin{tabular}{ c || r || r } \hline 
$n$ & $E^{B}_n/m \hspace{1cm} $ &$E^{F}_n/m \hspace{1cm} $  \\ \hline
 \ \  0 \ \ &\ \ 0.00000000001(3)\ \ & \\
 1 & \ 1.001498936(1) \    & \ \  1.00149893546(2)   \\
 2 & \ 2.00598024(3) \      & \ 2.005980230(1) \\
 3 & \ 3.0134265(5) \        & \ 3.01342635(3) \\
 4 & \ 4.023822(5) \          & \ 4.0238202(4) \\
 5 & \ 5.03716(4) \            & \ 5.037146(5)\\
 6 & \ 6.0535(3) \              & \ 6.05340(4)  \\
 7 & \ 7.073(1)  \               & \ 7.0726(2)\\
 8 & \ 8.097(5) \                & \ 8.095(1)\\
 9 & \ 9.13(2) \                  & \ 9.122(5)\\
 10 & \ 10.18(4) \              & \ 10.16(1)\\  \hline
\end{tabular}
\caption{Energy eigenvalues obtained from the  CLR action for $\lambda=0.001$ at $ma=0.01$.
}
\label{tab:energy_spectra_weak}
\end{center}
\end{table}

\begin{figure}[!htbp]
\includegraphics[]{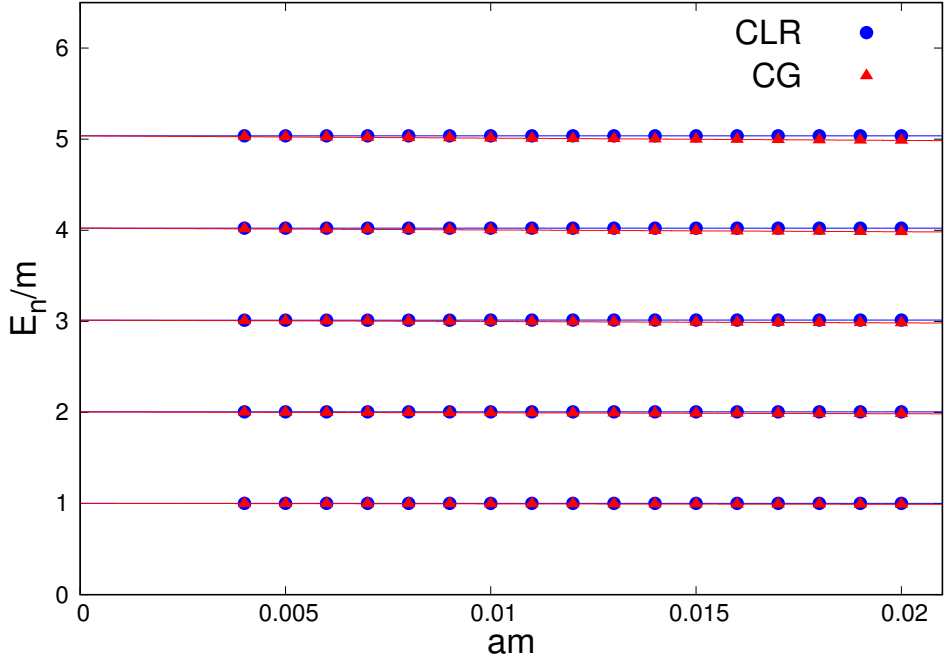}
\caption{Five lowest energy eigenvalues against the lattice spacing $ma$
             for $\lambda=0.001$. The results of CLR (circles)
             show a better convergence than the CG results (triangles). 
             The solid lines represent the fit results shown in
             Tables \ref{tab:CLR_energy_spectra_weak} and \ref{tab:CG_energy_spectra_weak}.
}
\label{fig:energy_spectra_weak}
\end{figure}

Figure \ref{fig:energy_spectra_weak} shows the lowest five eigenvalues against the lattice spacing $ma$.  
Since the difference between $E_n^B$ and $E_n^F$ are sufficiently smaller 
than the systematic errors from finite $K$ effect, we plotted only $E_n^F$ as $E_n$ in the figure.
As we can see, the cut-off dependence of the CLR action is milder than that of CG action.

Tables \ref{tab:CLR_energy_spectra_weak} and \ref{tab:CG_energy_spectra_weak} show 
the fit results of the lowest five energy eigenvalues for the CLR and CG actions, respectively.
For the continuum extrapolation, we employ a quadratic polynomial, 
\begin{equation}
\label{extrapolator}
E/m = a_0 + a_1(ma) + a_2 (ma)^2.
\end{equation}
Two actions reproduce the same $a_0$, which is $E_n/m$ at the continuum limit, 
within the errors.
The CLR action behaves rather similar to the continuum theory in comparison with the CG action 
as suggested from small values of $a_1$.

\begin{table}[!htbp]
\begin{center}
\begin{tabular}{ c  c  c  c  c  c} \hline \hline
           & $E_1/m$    & $E_2/m$ & $E_3/m$ & $E_4/m$ & $E_5/m$ \\ \hline \hline
 $a_0$ \hspace{1.5mm} & \hspace{3mm} 1.001495535(1) \hspace{3mm} & \hspace{3mm}2.00597334(3) \hspace{3mm} &\hspace{3mm} 3.0134165(4) \hspace{3mm} &\hspace{3mm} 4.023814(4) \hspace{3mm} & \hspace{3mm}5.0372(4) \\
 $a_1$ \hspace{1.5mm} & -0.0004999(4)    & -0.00101(1)      & -0.0017(2)      & -0.004(2)        & -0.02(2)   \\
 $a_2$ \hspace{1.5mm} & 0.08400(4)          & 0.1702(9)         &0.27(1)          & 0.5(2)           &  2(1)      \\ \hline \hline
\end{tabular}
\caption{Fit results of $E_n$ for the CLR action with $\lambda=0.001$.
}
\label{tab:CLR_energy_spectra_weak}
\end{center}
\end{table} 
\vspace{2mm}
\begin{table}[!htbp]
\begin{center}
\begin{tabular}{ c  c  c  c  c  c} \hline \hline
          & $E_1/m$    & $E_2/m$ & $E_3/m$ & $E_4/m$ & $E_5/m$ \\ \hline \hline
 $a_0$ \hspace{1.5mm} & \hspace{3mm}1.0014954(2) \hspace{3mm} &\hspace{3mm} 2.0059732(4) \hspace{3mm} &\hspace{3mm} 3.0134161(9) \hspace{3mm} &\hspace{3mm} 4.023806(1) \hspace{3mm} & \hspace{3mm} 5.037131(3) \\
 $a_1$ \hspace{1.5mm} & -0.50221(6)      & -1.0089(1)   & -1.5202(3)    & -2.0357(3)      & -2.557(1)        \\
 $a_2$ \hspace{1.5mm} & 0.330(3)        & 0.667(7)       &1.02(2)           &1.36(2)             & 1.8(1)             \\ \hline \hline
\end{tabular}
\caption{Fit results of $E_n$ for the CG action with $\lambda=0.001$.
}
\label{tab:CG_energy_spectra_weak}
\end{center}
\end{table}

The weak coupling expansion of the first excited energy is demonstrated in 
appendix \ref{sec:weak_coupling_expansion}, in which 
the quantum corrections to the masses are evaluated
from the correlation functions.
We  find that, for $E_1 \equiv E_1^F=E_1^B$, 
the one-loop result of the CLR action is 
\begin{equation}
\frac{E_1^{CLR}}{m} =1+\frac{3}{2}\lambda -\frac{1}{2}ma\lambda+O((ma)^2,\lambda^2),
\label{E1_PT_CLR}
\end{equation}
while one of the CG action is 
\begin{equation}
\frac{E_1^{CG}}{m} =1+\frac{3}{2}\lambda -\frac{ma}{2}-\frac{1}{2}ma\lambda+O((ma)^2,\lambda^3) .
\label{E1_PT_CG}
\end{equation}
Both one-loop results coincide with 
one of the continuum theory, $E^{cont}/m=1+\frac{3}{2}\lambda$, as $a \rightarrow 0$. 
The CG action has a large discretization error due to the third term of $O(ma)$ in (\ref{E1_PT_CG}), 
while the $O(a)$-term starts from $O(\lambda ma)$ in the CLR action, which is much smaller than $O(ma)$ for $\lambda=0.001$.

\begin{figure}[htbp]
\includegraphics[]{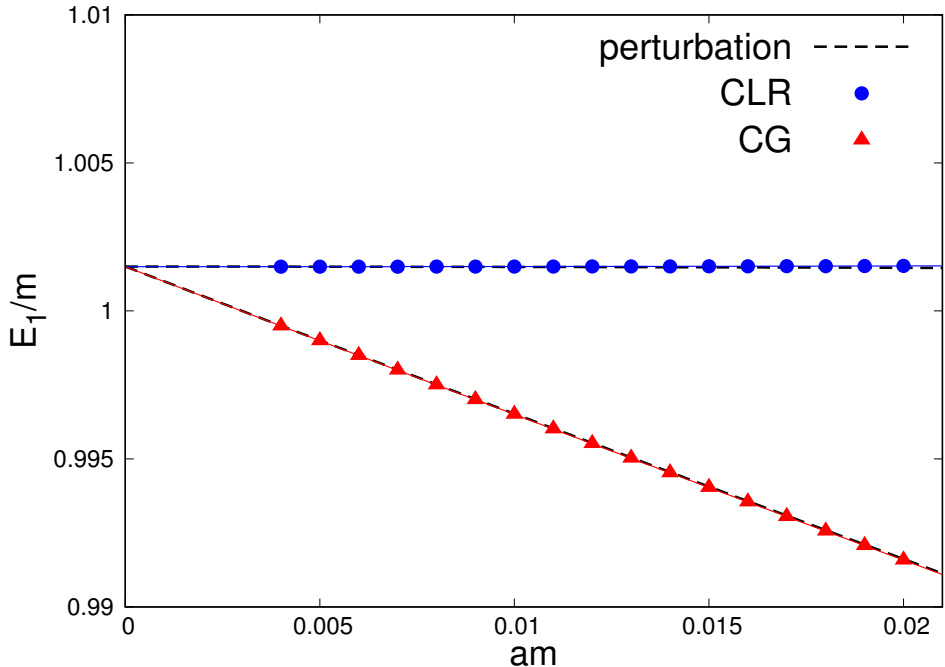}
\caption{Continuum limit of $E_1$ for $\lambda=0.001$. 
The solid lines represent the fit results and the dotted lines are the perturbative results}
\label{fig:e1_weak_coupling}
\end{figure}

Figure \ref{fig:e1_weak_coupling} shows the numerical results of
$E_1$ with the perturbative ones (\ref{E1_PT_CLR}) and  (\ref{E1_PT_CG}) for $ma \le 0.02$.
The numerical results nicely reproduce the perturbation theory shown by the dotted lines
and the relative errors are of the order of $10^{-6}$ that is the same size of $\lambda^2$.
Although a linear $ma$ dependence is seen in the CG-results, 
the CLR-results perfectly reproduce the continuum theory for this range of $ma$
since the third term of  (\ref{E1_PT_CLR}) is negligibly small for  $\lambda=0.001$.

\subsubsection{Strong coupling results}

Table \ref{tab:energy_spectra_strong} shows the ten smallest energy eigenvalues obtained from the CLR action 
for  $\lambda=1$ at a fixed $ma=0.01$. 
The central values are again ones evaluated for $K=150$ and 
the errors are estimated from the largest difference among the results for $K=140, 150, \cdots, 200$.
The energy spectra have large quantum corrections 
compared to Figure  \ref{tab:energy_spectra_weak} for the weak coupling $\lambda=0.001$.
$E_n^B$ and $E_n^F$  coincide with each other within the errors 
as well as the case of the weak coupling.

\begin{table}[!htbp]
\begin{center}
\begin{tabular}{ c || r || r} \hline 
$n$ & $E^{B}_n/m$ \hspace{0.8cm} &$E^{F}_n/m$  \hspace{0.8cm}  \\ \hline
\ \ 0 \ \  & \ \ 0.0000000000(2) \ & \\
 1 & \ 1.682687275(2) \ &\ \ 1.682687274859(4) \\
 2 & \ 4.365387624(8) \ & 4.36538762319(6)\ \\
 3 & \ 7.62211841(4)  \  & 7.6221184119(5)\ \\
 4 & \ 11.3640034(2)  \  & 11.364003389(3)\ \\
 5 & \ 15.5273615(7)  \  & 15.52736144(2)\ \\
 6 & \ 20.068372(3)    \  & 20.06837202(9)\  \\
 7 & \ 24.954588(9)    \  & 24.9545871(4)\ \\
 8 & \ 30.16073(3)      \  & 30.160725(2) \ \\
 9 & \ 35.66638(9)      \  & 35.666371(6) \ \\
 10 \ & 41.4546(3)      \  & 41.45459(2) \ \\  \hline
\end{tabular}
\caption{
Energy eigenvalues obtained from the  CLR action for $\lambda=1$ at $ma=0.01$.
}
\label{tab:energy_spectra_strong}
\end{center}
\end{table}

\begin{figure}[!htbp]
\includegraphics[]{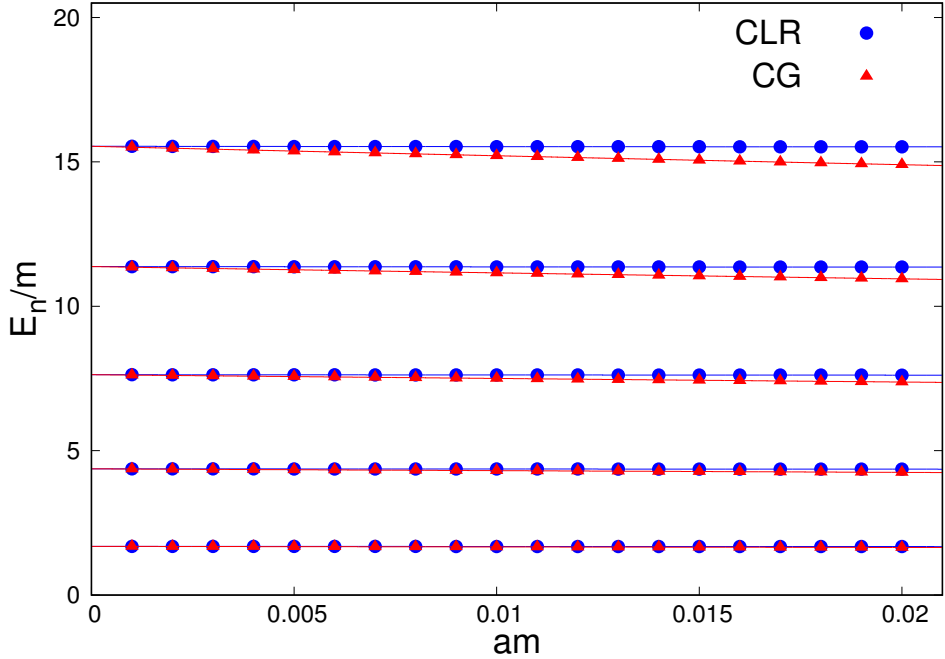}
\caption{
             Five lowest energy eigenstates against $ma$
             for $\lambda=1$. The results of CLR and CG are shown as the circles and 
             the triangles, respectively. 
             The solid lines represent the fit results shown in
             Tables \ref{tab:CLR_energy_spectra_strong} and \ref{tab:CG_energy_spectra_strong}.
             }
\label{fig:energy_spectra_strong}
\end{figure}

\begin{figure}[h]
\includegraphics[]{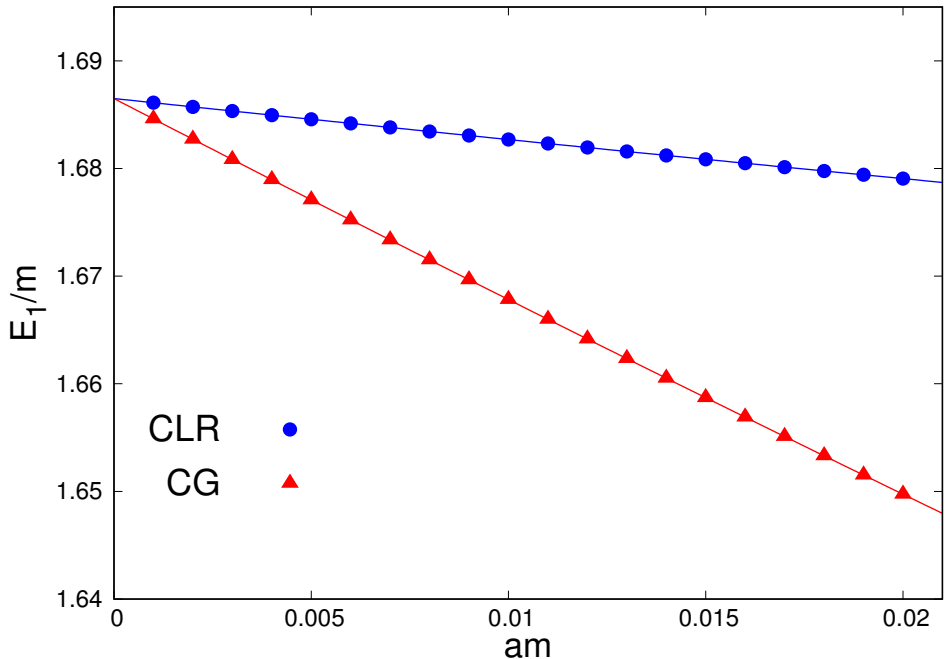}
\caption{Continuum limit of $E_1$ for $\lambda=1$.}
\label{fig:e1_strong_copling}
\end{figure}

Figure \ref{fig:energy_spectra_strong} shows the lowest five energy eigenvalues 
against $ma$ for $\lambda=1$. 
We also show Figure \ref{fig:e1_strong_copling} which focuses on $E_1$ for $\lambda=1$
for a comparison with Figure \ref{fig:e1_weak_coupling}.
The obtained $E_n^F$ is again plotted as $E_n$ since $E_n^F=E_n^B$ 
within the sufficiently small errors of $O(10^{-8})$.
The cut-off dependence of the CLR action is milder than that of CG action
as well as the weak coupling shown in Figure \ref{fig:energy_spectra_weak}.

Tables \ref{tab:CLR_energy_spectra_strong} and \ref{tab:CG_energy_spectra_strong} show the fit results 
of $E_n$ with a quadratic function (\ref{extrapolator}).
The same $a_0$ which is $E/m$ in the continuum limit are obtained between the CLR and CG actions.
As a visible difference between Figure \ref{fig:energy_spectra_weak} and 
Figure \ref{fig:energy_spectra_strong} can be seen,  
the coefficients $a_1$ and $a_2$ are  systematically larger than
those for the weak coupling, which are shown in  
Tables \ref{tab:CLR_energy_spectra_weak} and \ref{tab:CG_energy_spectra_weak}.
In the strong coupling region, we can confirm that 
the $O(a)$ dependence of $E_1$ obtained for the CLR action 
is still smaller than that of the CG action.

\begin{table}[!htbp]
\begin{center}
\begin{tabular}{ c  c  c  c  c  c} \hline \hline
           & $E_1/m$    & $E_2/m$ & $E_3/m$ & $E_4/m$ & $E_5/m$ \\ \hline \hline
 $a_0$ \hspace{1.5mm}& \hspace{3mm}1.6865004(6) \hspace{3mm}& \hspace{3mm} 4.371816(2) \hspace{3mm}&\hspace{3mm} 7.630953(5) \hspace{3mm}&\hspace{3mm} 11.374845(7) \hspace{3mm}& \hspace{3mm}15.53978(1) \\
 $a_1$ \hspace{1.5mm}& -0.3907(3)      & -0.684(1)      & -0.985(2)      & -1.282(3)        & -1.575(4)         \\
 $a_2$ \hspace{1.5mm}& 0.94(2)           & 4.08(8)         &10.2(2)          & 19.8(2)           &  33.3(3)             \\ \hline \hline
\end{tabular}
\caption{Fit results of $E_n$ for the CLR action with $\lambda=1$.
}
\label{tab:CLR_energy_spectra_strong}
\end{center}
\end{table}
\vspace{2mm}
\begin{table}[!htbp]
\begin{center}
\begin{tabular}{ c  c  c  c  c  c} \hline \hline
           & $E_1/m$    & $E_2/m$ & $E_3/m$ & $E_4/m$ & $E_5/m$ \\ \hline \hline
 $a_0$ \hspace{1.5mm}& \hspace{3mm} 1.686500(3) \hspace{3mm}& \hspace{3mm} 4.37181(1) \hspace{3mm}& \hspace{3mm} 7.63095(4) \hspace{3mm}& \hspace{3mm} 11.37483(8) \hspace{3mm}& \hspace{3mm} 15.5398(1) \\
 $a_1$ \hspace{1.5mm}& -1.898(1)      & -6.422(6)    & -13.30(2)    & -22.43(3)      & -33.75(6)         \\
 $a_2$ \hspace{1.5mm}& 3.05(9)         & 12.6(4)       &31(1)           & 58(5)           & 95(5)             \\ \hline \hline
\end{tabular}
\caption{ Fit results of $E_n$ for the CG action with $\lambda=1$.}
\label{tab:CG_energy_spectra_strong}
\end{center}
\end{table}

\subsection{SUSY WT identities}
\label{sec:susy_wti}

The CLR action has an exact SUSY parametrized by $\epsilon$ in (\ref{lattice_susy_CLR}) 
while the other $\bar\epsilon$ SUSY is broken at finite lattice spacing for any interacting case.
The correct mass spectra shown in the previous section imply that  
the broken $\bar\epsilon$ symmetry is restored in the continuum limit.
Testing the SUSY WTIs,
we study the restoration of the full SUSY.

To this end, we first define the SUSY WTIs on the lattice. 
The broken  $\bar\epsilon$ transformation cannot be uniquely defined on the lattice
because one can add several terms that vanish in the continuum limit to the transformation.
Here, for the CLR action, we employ (\ref{lattice_CLR_another}) 
as a lattice $\bar\epsilon$  transformation, 
which is an exact symmetry in the free theory. 
Correspondingly, we use (\ref{lattice_susy_CG}) for the CG action, 
whose $\bar\epsilon$-transformation is exactly kept in the free case of (\ref{S_CG}).

We can show that 
\begin{eqnarray}
\langle \delta(\phi_n \bar\psi_N + \psi_n \phi_N) \rangle  =  
\epsilon R_n + \bar\epsilon \bar R_n,
\end{eqnarray}
where 
\begin{eqnarray}
&& R_n\equiv \langle \psi_n\bar{\psi}_{N}\rangle 
- \langle \phi_n(\nabla_-\phi)_{N} \rangle 
-\langle\phi_{n}W_{N}\rangle, 
\label{WT1} \\
&&
\bar R_n \equiv  \langle \psi_n\bar{\psi}_{N}\rangle 
- \langle \phi_n(\nabla_-\phi)_{N} \rangle 
-\langle  W_{n+1}\phi_{N}\rangle, 
\label{WT2}
\end{eqnarray}
for the CLR action. For the CG action, $W_N$ and $W_{n+1}$ of (\ref{WT1}) and  (\ref{WT2}) 
are replaced by $W(\phi_N)$ and $W(\phi_n)$, respectively. 
The second term of $\bar R_n$ is actually found as $\langle (\nabla_+\phi)_n \phi_N \rangle $ 
which can be written as the same form as the second term of $R_n$ using the translational invariance.
Note that the third term is the only difference between $R_n$ and $\bar R_n$.

For any interacting case, we have  $R_n=0$  
since the $\epsilon$-transformation is an exact symmetry of the lattice actions.
However, $\bar R_n$ does not vanish at any finite lattice spacing for the interacting cases 
even if it vanishes for the  free theory. If the $\bar\epsilon$-symmetry is restored 
at a quantum continuum limit, 
$\bar R_n$ should approach zero as $a \rightarrow 0$.  
We evaluate  $\bar R_n$ numerically 
to confirm whether the second SUSY WTI is restored in the continuum limit or not,
as already done for the CG action in Ref.\cite{Bergner:2007pu}.

\begin{figure}[!htbp]
\includegraphics[]{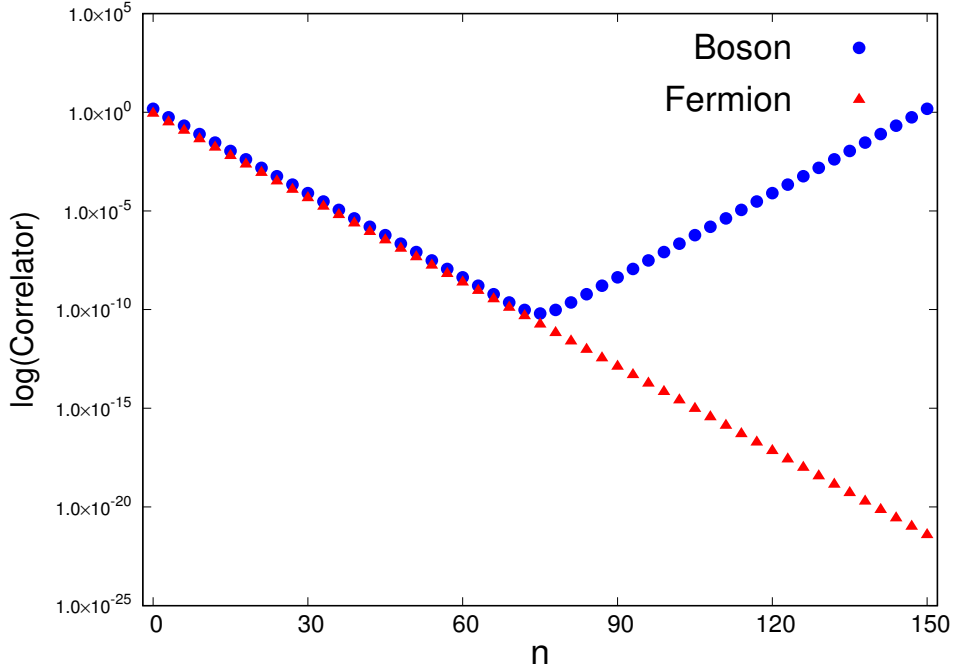}
\caption{ $\langle \phi_n \phi_N \rangle$ and  $\langle \psi_n \bar{\psi}_N \rangle$ obtained from the CLR action 
             for $\lambda=1$ and $ma=0.2$. 
             The $x$-axis denotes the lattice site $n$ and the $y$-axis shows the numerical values of the correlators
             in the logarithmic scale.}
\label{fig:correlation_function}
\end{figure}

Figure \ref{fig:correlation_function} shows $\langle \phi_n \phi_N \rangle$
and $\langle \psi_n \bar{\psi}_N \rangle$ for $\lambda=1$ and $ma=0.2$.
When $N$ is sufficiently large, 
as confirmed in the figure,
$\langle \phi_n \phi_N \rangle$ and  $\langle \psi_n \bar{\psi}_N \rangle$  behave as 
\begin{eqnarray}
&&\langle \phi_n \phi_N \rangle \approx 
C (e^{- an E_1 } + e^{- a(N- n) E_1 }),\\
&&\langle \psi_n \bar{\psi}_N \rangle \approx  D e^{- an E_1},
\label{psipsi}
\end{eqnarray}
for  $1  \ll n\ll N$. Here $C$ and $D$ are some constants 
that depend on the lattice spacing. 
Similarly, using the translational invariance, 
the other correlation functions in $R_n$ and $\bar R_n$ 
are expected to be
\begin{eqnarray}
&&\langle  \phi_n \nabla_- \phi_N \rangle \approx 
C_1 (e^{- an E_1 } -  e^{- a(N- n-1) E_1 }),
\label{dpp}
\\  
&& \langle \phi_{n} W_N \rangle \approx 
C_2 e^{- an E_1 } + C_3 e^{- a(N- n-1) E_1 },
\label{pW}\\
&& \langle W_{n+1} \phi_N \rangle \approx 
C_3 e^{- an E_1 } +   C_2 e^{- a(N- n-1) E_1 },
\label{pW2}
\end{eqnarray}
for $1 \ll n\ll N$. 
Here $C_1=C(1-e^{-a E_1})/a$ and $C_2, C_3$ are some constants 
that depend on the lattice spacing. 
Note that it is possible to ignore the contribution 
from the second excited state for $1 \ll n\ll N$. 
We can immediately show that
\begin{eqnarray}
C_1= C_3 = D-C_2  
\end{eqnarray}
from $R_n = 0$ and  
the second WTI holds if and only if $C_2 \rightarrow C_3$ as $a \rightarrow 0$.

\begin{figure}[!htbp]
  \begin{center}
    \begin{tabular}{c}
      \begin{minipage}{0.5\hsize}
        \begin{center}
          \includegraphics[clip, width=8.5cm]{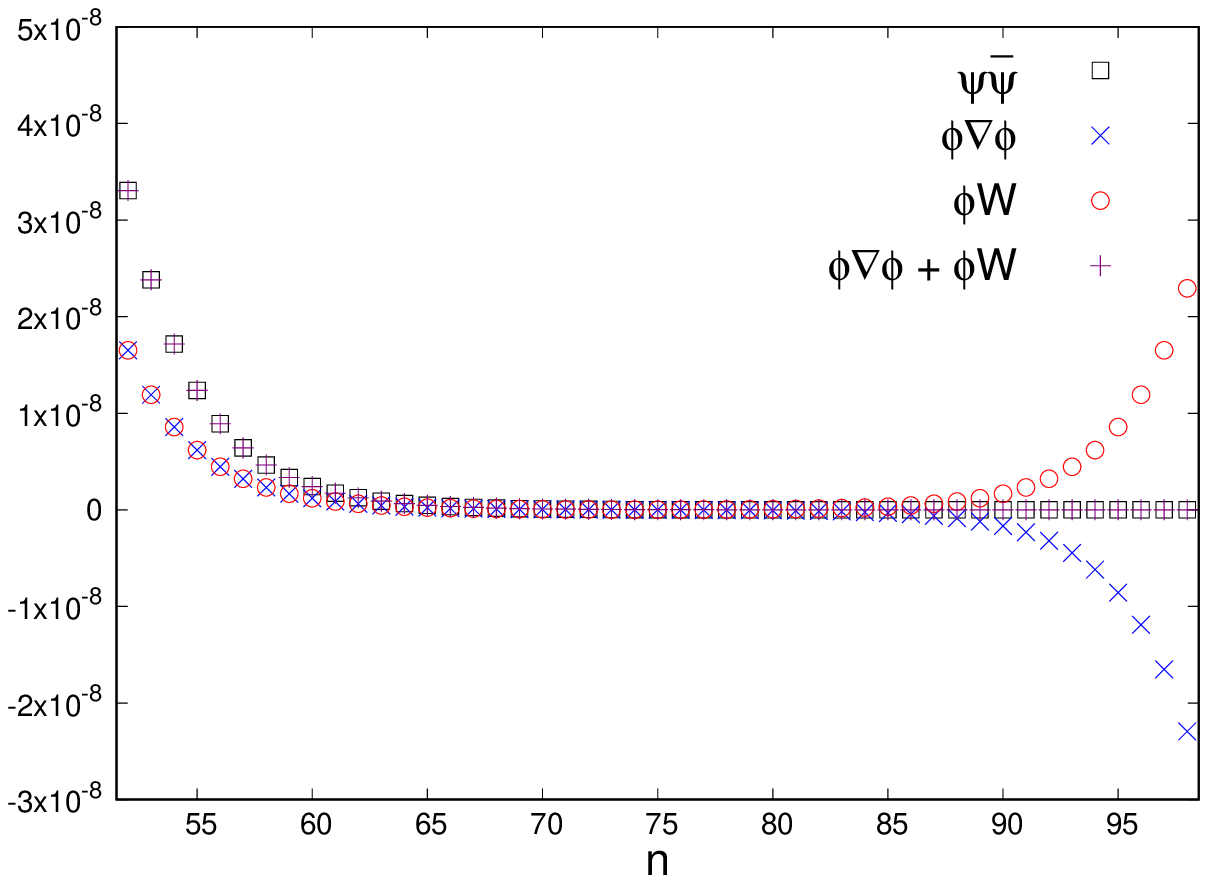}
        \end{center}
      \end{minipage}
      
      \begin{minipage}{0.5\hsize}
        \begin{center}
          \includegraphics[clip, width=8.5cm]{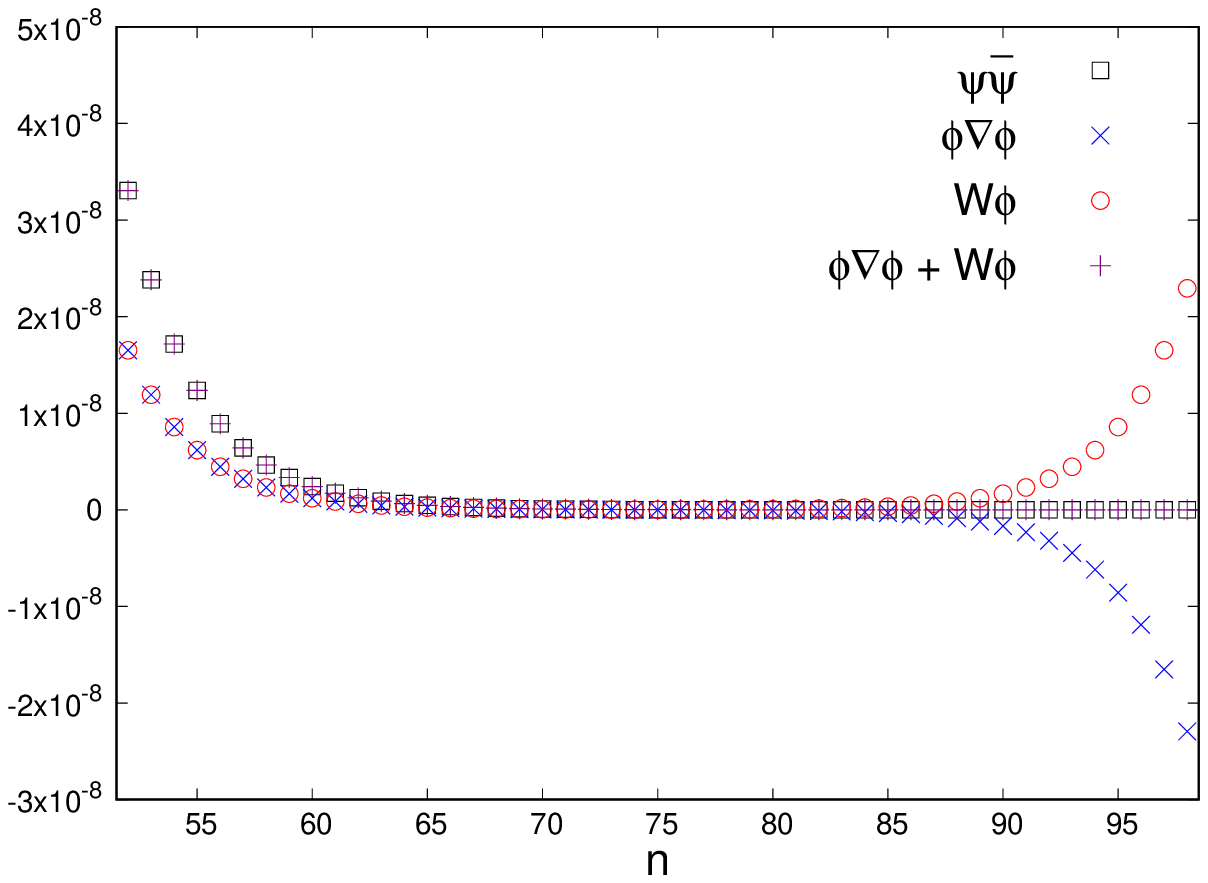}
        \end{center}
      \end{minipage}
      
    \end{tabular}
    \caption{Three correlation functions in ${R}_n$ (left) and $\bar {R}_n$ (right) for the CLR action. 
     The cancellations among them are clearly observed. }
    \label{fig:CLRcancellation}
  \end{center}
\end{figure}

\begin{figure}[!htbp]
  \begin{center}
    \begin{tabular}{c}
      \begin{minipage}{0.5\hsize}
        \begin{center}
          \includegraphics[clip, width=8.5cm]{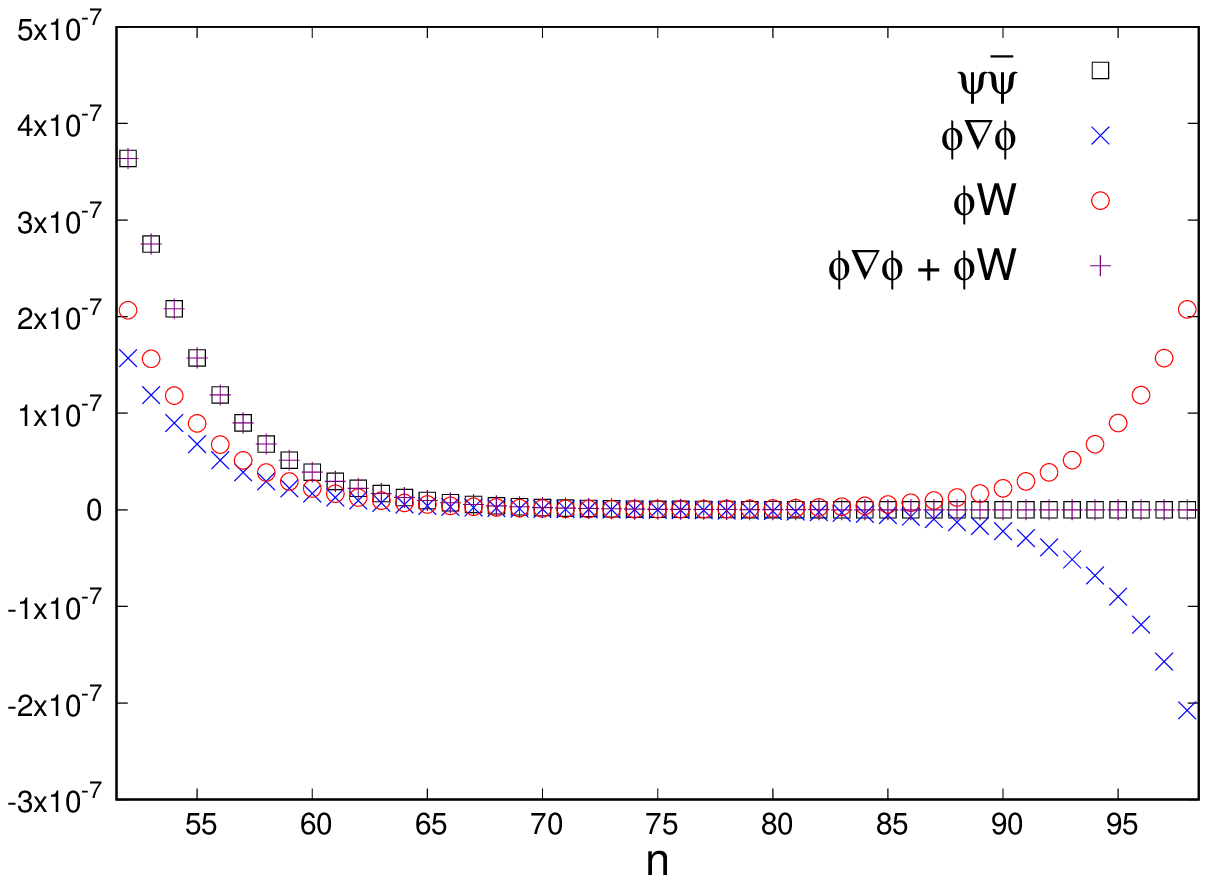}
        \end{center}
      \end{minipage}
      
      \begin{minipage}{0.5\hsize}
        \begin{center}
          \includegraphics[clip, width=8.5cm]{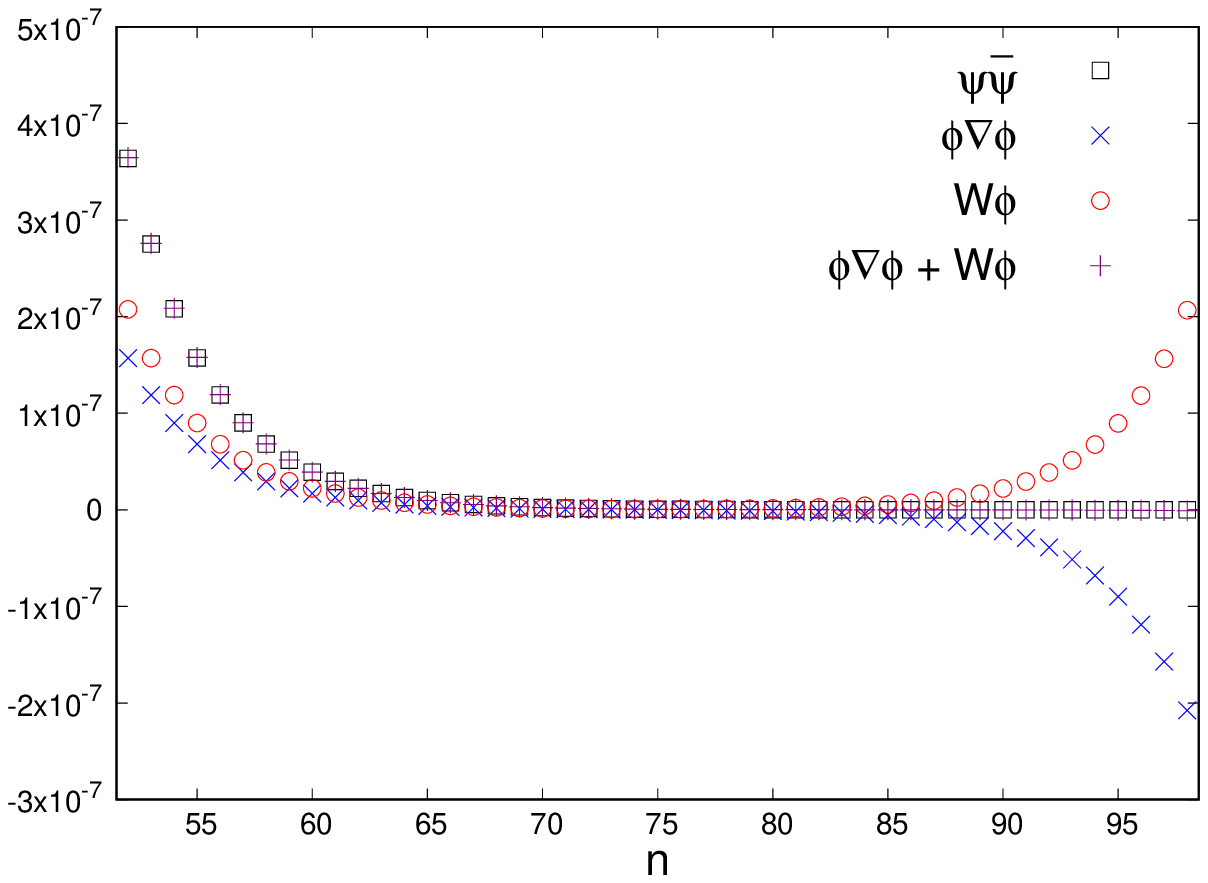}
        \end{center}
      \end{minipage}
      
    \end{tabular}
    \caption{
      Three correlation functions in ${R}_n$ (left) and $\bar {R}_n$ (right) for the CG action. 
     The cancellations are observed as well as the CLR 
      case shown in Figure \ref{fig:CLRcancellation}. }
    \label{fig:CGcancellation}
  \end{center}
\end{figure}

Figure \ref{fig:CLRcancellation} shows 
the cancellation among three correlation functions in $R_n$ (Left) and $\bar{R}_n$ (Right) 
for $\lambda=1$ and $ma=0.2$. 
In Figure \ref{fig:CGcancellation}, the similar plots obtained for the CG action are shown. 
As we expected, the other correlators 
show the behavior of (\ref{dpp}),  (\ref{pW}) and (\ref{pW2}).
The cancellation for $n<N/2$ is realized in a different way from 
that of $n>N/2$. 
As suggested from (\ref{psipsi})-(\ref{pW2}), the sum of two bosonic correlators (denoted as crosses) 
cancels the fermion correlator (denoted as squares)  
for $1 \ll n \ll N/2$
while  two bosonic correlators cancel each other out for $N/2 \ll n \ll N$
since the fermion correlator is approximately zero compared 
with the others. 

Since each term of $R_n$ and $\bar R_n$ is very small for $n \simeq N/2$, 
we normalize them to observe 
the effect of the breaking term clearly:  
\begin{eqnarray}
&& S_n \equiv \frac{R_n}{|\langle \psi_n\bar{\psi}_{N}\rangle|
+|\langle \phi_n (\nabla_-\phi)_{N} \rangle|+|\langle\phi_{n}W_{N}\rangle|}
\label{NWT1} 
\\
&&\bar{S}_n \equiv \frac{\bar R_n}{|\langle \psi_n\bar{\psi}_{N}\rangle|
+|\langle \phi_n (\nabla_- \phi)_N \rangle |+|\langle  W_{n+1}\phi_{N}\rangle |}.
\label{NWT2}
\end{eqnarray}
Note again that $W_N$ and $W_{n+1}$ of (\ref{NWT1}) and  (\ref{NWT2}) 
are replaced by $W(\phi_N)$ and $W(\phi_n)$, respectively, for the CG action.
It is immediately found that $S_n=0$ for any $n$ since $R_n=0$.

The asymptotic behavior of $\bar S_n$ 
can be understood from (\ref{psipsi}), (\ref{dpp}) and (\ref{pW2}). 
For sufficiently large $N$, it can be shown that $\bar S_n$ behaves 
as constants: 
\begin{eqnarray}
S_n \approx h_1 \equiv \frac{C_2-C_3}{2 |C_3|+ |D|},  & \qquad  (1 \ll n \ll N/2)
\label{h1}
\end{eqnarray}
and 
\begin{eqnarray}
S_n \approx h_2 \equiv \frac{C_3-C_2}{ |C_2|+|C_3|},  & \qquad  (N/2 \ll n \ll N).
\label{h2}
\end{eqnarray}
We have 
\begin{eqnarray}
h_2 = -2 h_1 +O(h_1^2), 
\label{h1h2}
\end{eqnarray}
when $C_2$ and $C_3$ have the same sign. 
The similar identities as (\ref{h1}),  (\ref{h2}) and  (\ref{h1h2})  hold for the CG action.

\begin{figure}[!htbp]
   \begin{center}       
          \includegraphics[]{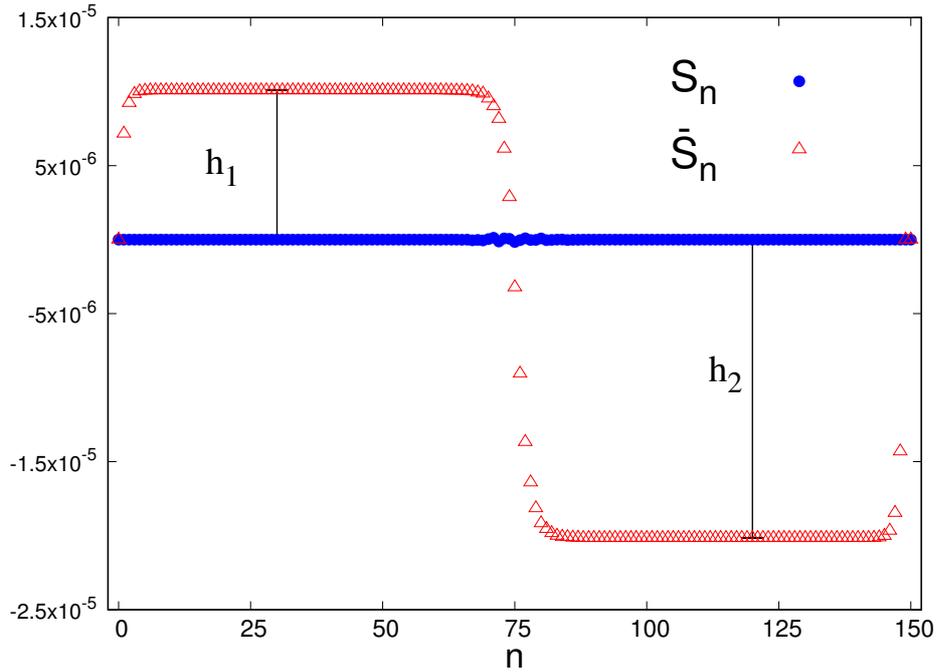}
         \caption{$S_n$ and $\bar S_n$ for the CLR action with $\lambda=1$ and $ma=0.2$.}
    \label{fig:WT_id_CLR}
  \end{center}
\end{figure}

\begin{figure}[!htbp]
  \begin{center}
           \includegraphics[]{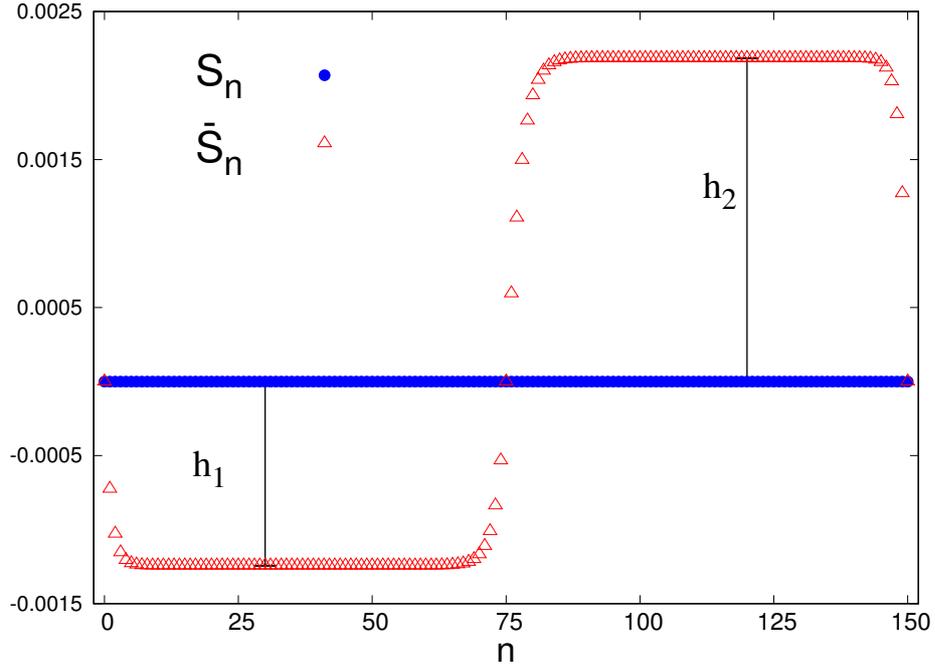}
           \caption{$S_n$ and $\bar S_n$ for the CG action with $\lambda=1$ and $ma=0.2$.}
           \label{fig:WT_id_CG}
   \end{center}         
\end{figure}

In Figure \ref{fig:WT_id_CLR} and Figure \ref{fig:WT_id_CG},  ${S}_n$  and 
$\bar{S}_n$ are plotted against $n$. As we expected, 
$S_n$ vanishes as numerical results while  
 $\bar{S}_n$ has two  plateaux 
corresponding to $h_1$ and $h_2$. 
We should note that the scale of the $y$-axis 
for the CLR action is rather smaller than that of the CG action.  
The value of $\bar S_n$ rapidly changes from $h_1$ to $h_2$ 
around  $n=N/2$ as a result of the cancellation of three correlation functions.  

\begin{figure}[!htbp]
 \begin{center}
  \includegraphics[]{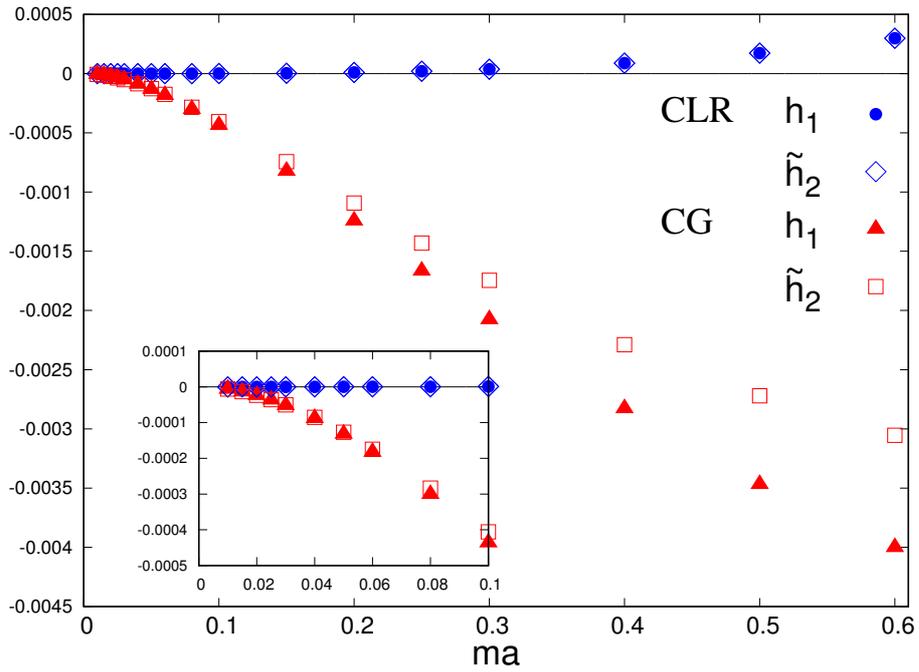}
  \caption{Lattice spacing dependence of $h_1$ and $h_2$ for $\lambda=1$.  
  We plot $h_1$ and $\tilde{h}_2=-h_2/2$, which are evaluated at $n = N/5$ and $n = 4N/5$,
  as circles and diamonds for the CLR action 
   and triangles and squares for the CG action. }
  \label{fig:comparison_h_strong}
 \end{center}
\end{figure}

\begin{figure}[!htbp]
 \begin{center}
  \includegraphics[]{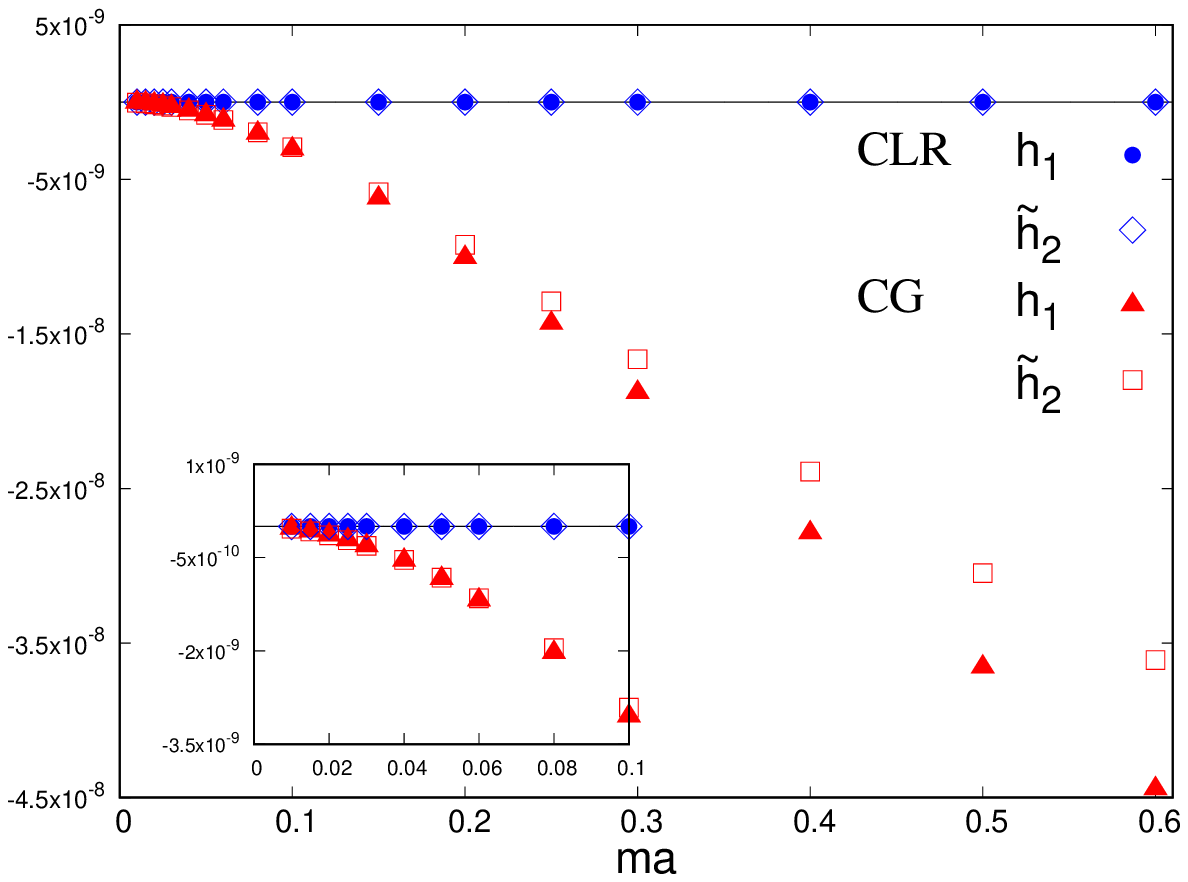}
  \caption{Lattice spacing dependence of $h_1$ and $h_2$ for $\lambda=0.001$.  
  We plot $h_1$ and $\tilde{h}_2=-h_2/2$, which are evaluated at $n = N/5$ and $n = 4N/5$,
   as circles and diamonds for the CLR action 
   and triangles and squares for the CG action. }
  \label{fig:comparison_h_weak}
 \end{center}
\end{figure}

Figures \ref{fig:comparison_h_strong} shows 
the lattice spacing dependence of $h_1$ and $h_2$ for $\lambda=1$ 
and the numerical values are shown in Table \ref{tab:h1_and_h2}
for the convenience of further studies.
Figure \ref{fig:comparison_h_weak} shows the same plot for $\lambda=0.001$.
We evaluate $h_1$ and $h_2$ at $n= N/5$ and $n=4N/5$,  respectively.
It can be seen that $h_1$ and $h_2$ approach zero as $a \rightarrow 0$.   
Consequently,  the second SUSY WTI holds in the continuum limit, that is, 
full SUSY is restored in the quantum continuum limit 
at low energy region $1 \ll n \ll N$. The breaking effect $h_1$ and $h_2$ 
of the CLR action are significantly smaller 
than the  CG action even for the strong coupling.
Thus we can conclude that the CLR shows a good behavior that is similar to the continuum theory 
at a non-perturbative level.

\begin{table}[!bhtp]
\begin{center}
\footnotesize
\begin{tabular}{ c | r | r || r | r } \hline 
 &  \multicolumn{2}{c||}{\normalsize{CLR}} & \multicolumn{2}{c}{\normalsize{CG}}  \\ \hline
\normalsize{$ma$}& \normalsize{$h_1$} \hspace{1.5cm} & \normalsize{$h_2$} \hspace{1.5cm} &\normalsize{$h_1$} \hspace{1.5cm} &\normalsize{$h_2$} \hspace{1.5cm} \\ \hline
 0.600 \  & \ \ \ 2.99391782(7)$\times10^{-4} $  \ & \ \ \    -5.9860435(2)$\times10^{-4}$ \ & \ \ \  -4.0012296(2)$\times10^{-3}$ \ & \ \ \ 6.10759335(1)$\times10^{-3}$ \\
 0.500 \ & \ 1.727581(1)$\times10^{-4}$ \ & \  -3.454565(2)$\times10^{-4}$ \ & \ -3.4686833(2)$\times10^{-3}$ \ & \ 5.44024340(3)$\times10^{-3}$ \\
 0.400 \ & \ 8.8385(4)$\times10^{-5}$ \ & \  -1.7675(1)$\times10^{-4}$ \ & \ -2.8283733(2)$\times10^{-3}$ \ & \ 4.57824926(3)$\times10^{-3}$\\
 0.300 \ & \ 3.66716(2)$\times10^{-5}$ \ & \  -7.33404(4)$\times10^{-5}$ \ & \ -2.0775005(1)$\times10^{-3}$ \ & \ 3.4908132(8)$\times10^{-3}$ \\
 0.250 \ & \ 2.068870(6)$\times10^{-5}$ \ & \ -4.13765(1)$\times10^{-5}$ \ & \ -1.6669089(1)$\times10^{-3}$ \ & \ 2.8615897(1)$\times10^{-3}$\\
 0.200 \ & \ 1.008226(2)$\times10^{-5}$ \ & \  -2.01643(1)$\times10^{-5}$ \ & \ -1.24342044(8)$\times10^{-3}$ \ & \ 2.1858056(1)$\times10^{-3}$ \\
 0.150 \ & \ 3.87190(2)$\times10^{-6}$ \ & \ -7.7438(4)$\times10^{-6}$ \ & \ -8.2362334(4)$\times10^{-4}$ \ & \ 1.4866210(1)$\times10^{-3}$\\
 0.100 \ & \ 9.51290(6)$\times10^{-7}$ \ & \ -1.90258(8)$\times10^{-6}$ \ & \ -4.3652327(1)$\times10^{-4}$ \ & \ 8.116901(1)$\times10^{-4}$\\
 0.080 \ & \ 4.2875(1)$\times10^{-7}$ \ & \ -8.575(1)$\times10^{-7}$ \ & \ -3.0105156(6)$\times10^{-4}$ \ & \ 5.6710394(7)$\times10^{-4}$\\
 0.060 \ & \ 1.50127(2)$\times10^{-7}$ \ & \ -3.003(1)$\times10^{-7}$ \ & \ -1.8301692(4)$\times10^{-4}$ \ & \ 3.495123(2)$\times10^{-4}$\\
 0.050 \ & \ 7.6345(2)$\times10^{-8}$ \ & \ -1.527(1)$\times10^{-7}$ \ & \ -1.3229006(2)$\times10^{-4}$ \ & \ 2.5444813(8)$\times10^{-4}$\\  
 0.040 \ & \ 3.3033(3)$\times10^{-8}$ \ & \ -6.607(1)$\times10^{-8}$ \ & \ -8.8205996(4)$\times10^{-5}$ \ & \ 1.709075(1)$\times10^{-4}$\\
 0.030 \ & \ 1.1062(6)$\times10^{-8}$ \ & \ -2.21(2)$\times10^{-8}$ \ & \ -5.1741110(6)$\times10^{-5}$ \ & \ 1.010147(1)$\times10^{-4}$\\
 0.025 \ & \ 5.49(2)$\times10^{-9}$ \ & \ -1.10(4)$\times10^{-8}$ \ & \ -3.6707175(4)$\times10^{-5}$ \ & \ 7.19415(1)$\times10^{-5}$ \\
 0.020 \ & \ 2.318(6)$\times10^{-9}$ \ & \ -4.6(1)$\times10^{-9}$ \ & \ -2.4006488(8)$\times10^{-5}$ \ & \ 4.72349(1)$\times10^{-5}$ \\
 0.015 \ & \ 7.56(4)$\times10^{-10}$ \ & \ -1.5(4)$\times10^{-9}$ \ & \ -1.3803022(8)$\times10^{-5}$ \ & \ 2.72672(4)$\times10^{-5}$\\
 0.010 \ & \ 1.5(1)$\times10^{-10}$ \ & \ -3(1)$\times10^{-10}$ \ & \ -6.27257(1)$\times10^{-6}$ \ & \ 1.24415(2)$\times10^{-5}$\\ \hline
\end{tabular}
\caption{Numerical values of $h_1$  and $h_2$ for $\lambda=1$.}
\label{tab:h1_and_h2}
\end{center}
\end{table}

%% file: sec5.tex
\section{Summary and discussion}
\label{sec:summary}

The property of 
the cyclic Leibniz rule has been studied 
in ${\cal N}=2$ SUSY QM beyond the perturbation theory.
We have defined the lattice action on the basis of the CLR with the backward difference 
operator giving a solution for any superpotential.  
The numerical computations 
have been carried out using the transfer matrix  representation of the partition function 
and the correlation functions. 
Tuning the rescaling parameter, 
the energy spectra and SUSY Ward-Takahashi identities are obtained in a high accuracy. 
We have compared them with those of the Catterall-Gregory action.

Although the number of exact symmetry is the same between the CLR and the CG actions,
the CLR action provides a milder cut-off dependence of energy spectra 
for both weak and strong couplings. 
In the weak coupling limit, 
the ${\cal O}(a)$ term does not appear in the energy spectra for the CLR action
 but does for the CG action. 
Even for the strong coupling, we have observed that the coefficient of ${\cal O}(a)$ term for the CLR action
is smaller than the CG action. 
The lattice SUSY WTIs have shown the same tendency in the cut-off behavior.

In the ${\cal N}=4$ case with the CLR,
the number of exact SUSY is greater than 
the other lattice formulation.
We can expect that a lattice theory with the CLR is highly improved and behaves much similar to the continuum theory.
The results shown in this paper could be useful to construct the SUSY action 
with a modified Leibniz rule in higher dimensions.

%% file: sec_app.tex
\section{More about the CLR}
\label{more_about_CLR}

\subsection{Solutions for other difference operators}

The solution of the CLR for the forward difference operator 
and a symmetric difference operator $\nabla_S=\frac{1}{2} (\nabla_+-\nabla_-)$
are presented. 
For any difference operator $\nabla$,
the CLR is defined in the same manner as (\ref{general_CLR}). 
By repeating the same procedures as in section \ref{sec:CLR_solution}, 
we find that  (\ref{general_CLR}) can be written as
\begin{equation}
\sum_n \nabla \phi_n W_n =0. 
\label{CLR3_any_nabla}
\end{equation}
It is then easy to find a local solution of (\ref{CLR3_any_nabla}):
\begin{eqnarray}
W_{n}
= 
\left\{
\begin{array}{cc}
\vspace{-7mm}  \\
\frac{U(\phi_{n+1})-U(\phi_{n})}{\phi_{n+1}-\phi_{n}}   &    \quad {\rm for} \ \  \nabla= \nabla_+  \vspace{3mm} \\
 \frac{U(\phi_{n+1})-U(\phi_{n-1})}{\phi_{n+1}-\phi_{n-1}} &  \quad {\rm for} \ \  \nabla= \nabla_S
\vspace{1mm}  \\
\end{array}
\right.
\end{eqnarray}
The same discussions as mentioned in section \ref{sec:CLR_solution} 
tell us that $W_{n}$ is  well-defined local function that reproduces $W(\phi_n)$ up to ${\cal O}(a)$.

\subsection{The $m$-body CLR}
\label{mbodyCLR}

We now consider $W(\phi)=\sum_{m=0}^\infty c_m \phi^m$ with coupling constants $c_m$.
Then the lattice superpotential  $W_n$ is also expressed as a expansion,  
\begin{eqnarray}
W_n \equiv  \sum_{\ell=0}^{\infty} c_\ell [\phi]^\ell_n
\end{eqnarray}
with 
\footnote{
The simplest example of $M$ (but it is not a solution of CLR) is  
$M_{n,m_1,m_2, \ldots,m_\ell} = \delta_{nm_1} \delta_{nm_2} \cdots \delta_{nm_\ell}$.
Then the lattice action  (\ref{CLR_action}) coincides with the naive one 
owing to $W_n=W(\phi_n)$ and $W^\prime_{nm}=W^\prime(\phi_n) \delta_{nm}$.
We can express a scattering of lattice variables around the site $n$ 
by $M_{n,m_1,m_2, \cdots,m_\ell}$.
}
\begin{eqnarray}
[\phi]^\ell_n \equiv \sum_{m_1,m_2, \cdots,m_\ell} 
M_{n,m_1,m_2, \ldots,m_\ell} \phi_{m_1} \phi_{m_2} \cdots \phi_{m_\ell},
\label{general_product}
\end{eqnarray}
Here we assume that $M_{n,m_1,m_2, \cdots,m_\ell}$ is totally symmetric for $m_1,m_2,\ldots,m_\ell$ 
except for the first index $n$
and $[1]^\ell_n=1$ as an overall normalization. 
The locality condition is strictly defined as 
\begin{eqnarray}
|M_{n,m_1,m_2, \cdots,m_\ell}| < 
C {\rm exp}\{-\rho |n-m_k|\},
\label{locality_M}
\end{eqnarray}
where $C$ and $\rho>0$ are some positive constants for $k=1,\ldots, \ell$.
The summation in (\ref{general_product}) is well-defined because it is absolutely convergent 
for (\ref{locality_M}).

The CLR  in (\ref{CLR2}) is shown to be  
\begin{eqnarray}
&& \sum_{n}\Big\{\nabla_{nk} M_{n,n_1,n_2,\cdots,n_{m-1}, n_m}+\nabla_{n n_1} M_{n,n_2,n_3,\cdots, n_m,k} + \nonumber \\
&& \hspace{2cm} \cdots +
\nabla_{nn_m} M_{n, k, n_1,\cdots,n_{m-2}, n_{m-1}}
\Big\} =0,
\label{CLR1}
\end{eqnarray}
which is referred to as $m$-body CLR.
It is easy to show that
(\ref{CLR1}) is equivalent to (\ref{CLR2}).   
We should note that the indices $k,n_1,n_2,\cdots,n_m$ cyclically appear in (\ref{CLR1}).
This is the reason why we called (\ref{CLR2}) the {\it cyclic} Leibniz rule.

The solutions of the $m$-body CLR for the backward difference operator
can be read from (\ref{generic-potential}) using 
\begin{eqnarray}
M_{n,m_1,m_2,\cdots, m_{\ell}} = \frac{1}{\ell!} 
\frac{\partial^\ell  W_n}{\partial \phi_{m_1}\partial \phi_{m_2} \cdots \partial \phi_{m_\ell}} \bigg|_{c_m=1,\phi=0}.
\end{eqnarray}
We have
\begin{eqnarray}
&& M_{n,m}= \frac{1}{2} \left(\delta_{nm} + \delta_{n-1,m} \right), 
\label{solution_1body_CLR}
\\
&& M_{n,m,k}= 
\frac{1}{6} \left(
2 \delta_{nm} \delta_{nk} 
+ \delta_{n-1,m} \delta_{nk} 
+ \delta_{nm} \delta_{n-1,k} 
+ 2 \delta_{n-1,m} \delta_{n-1,k}
\right), 
\label{solution_2body_CLR}
\\
&& M_{n,m,k,l}=\frac{1}{12}( 3\delta_{n,m}\delta_{nk}\delta_{nl} 
+\delta_{n-1,m}\delta_{nk}\delta_{nl}
+\delta_{nm}\delta_{n,k+1}\delta_{nl}
+\delta_{nm}\delta_{nk}\delta_{n-1,l}
\nonumber \\
&&\hspace{1cm} 
+\delta_{n-1,m}\delta_{n-1,k}\delta_{nl}
+\delta_{n-1,m}\delta_{nk}\delta_{n-1,l} 
+\delta_{nm}\delta_{n-1,k}\delta_{n-1,l} 
+3\delta_{n-1,m}\delta_{n-1,k}\delta_{n-1,l}
), 
\label{solution_3body_CLR}
\end{eqnarray}
and so on.

The explicit forms of $M_{n,m_1,m_2, \cdots,m_\ell}$ for the forward difference operator  
are ones obtained by replacing the lattice site $n-1$  by $n+1$ in 
(\ref{solution_1body_CLR}), (\ref{solution_2body_CLR}) 
and (\ref{solution_3body_CLR}). Those for the symmetric difference operator 
$\nabla_S=\frac{1}{2}(\nabla_++\nabla_-)$
are also obtained by the similar replacement of the lattice site.

\section{Weak coupling expansion}
\label{sec:weak_coupling_expansion}

The weak coupling expansion of the first excited energy are presented at one-loop order
for the naive, the CG and the CLR actions.
We perform the lattice perturbation theory on the infinite volume lattice. 
The first excited energy are evaluated as effective masses 
obtained from the two-point correlation functions. 
In this section,  we assume $m>0$ and basically take $a=1$ 
except for  final results of the effective masses.

\subsection{Perturbative calculation on the infinite volume lattice}

The free part of a lattice action $S$ can be expressed 
in the momentum space as
\begin{eqnarray}
S_{\rm free} =  \int^{\pi}_{-\pi} \frac{dp}{2\pi}  \left\{
\frac{1}{2} D_0^{-1}(p) \phi(p) \phi(-p)  + S_0^{-1}(p) \bar\psi(p) \psi(-p) 
\right\},
\end{eqnarray} 
where $D_0(p)$ and $S_0(p)$ are bare propagators of the boson and the fermion, respectively.
The concrete form of $D_0(p)$ and $S_0(p)$, which depends on $S_{\rm free}$, are obtained by  
the Fourier transformation for a lattice variable $\varphi_n$:  
\begin{eqnarray}
&& \varphi (p) = \sum_{n \in \mathbb{Z}}  {\rm e}^{ipn} \varphi_n, \\
&& \varphi_n = \int^{\pi}_{-\pi} \frac{dp}{2\pi}  {\rm e}^{-ipn} \varphi (p), 
\end{eqnarray}
with a useful identity $\delta_{n0} = \int^{\pi}_{-\pi} \frac{dp}{2\pi}  {\rm e}^{ipn} \, (n \in \mathbb{Z})$.
Note that $\varphi (p + 2\pi m) =\varphi(p)$ for $m \in \mathbb{Z}$.

The two-point correlation functions are defined as
\begin{eqnarray}
&&D_{kl} \equiv \langle  \phi_k  \phi_l \rangle  =\int_{-\pi}^{\pi} \frac{dp}{2\pi} D(p) e^{ip(k-l)},
\label{boson_propagator}
\\
&& S_{kl} \equiv \langle  \psi_k  \bar{\psi}_l \rangle  = \int_{-\pi}^{\pi} \frac{dp}{2\pi} S (p)e^{ip(k-l)},
\label{fermion_propagator}
\end{eqnarray}
where $D(p)$ and $S(p)$ are the full propagators. 
We have $D_{mn}=D_{m-n,0}$ and $S_{mn}=S_{m-n,0}$ as a result of the translational invariance. 
The free two-point correlation functions $(D_0)_{kl}$ and  $(S_0)_{kl}$ are calculated 
from (\ref{boson_propagator}) and  (\ref{fermion_propagator}) with $D_0(p)$ and $S_0(p)$  using 
the complex integral with $z=e^{ip}$.

The full propagators can be evaluated 
in the weak coupling expansion from $D_0, S_0$ and the boson and the fermion 
self energies $\Pi_{kl}$ and $\Sigma_{kl}$. 
As well-known,  
$D_{kl}$ is given by an infinite series,  
\begin{eqnarray}
D_{kl} = D_{0,kl} - (D_0 \Pi D_0)_{kl}  + (D_0 \Pi D_0 \Pi D_0)_{kl} - \ldots.
\end{eqnarray}
Thus we have
\begin{eqnarray}
D_{kl} = \left(\frac{1}{D_0^{-1} + \Pi}\right)_{kl}.
\label{full_D}
\end{eqnarray}
Similarly, 
\begin{eqnarray}
S_{kl} = \left(\frac{1}{S_0^{-1} + \Sigma}\right)_{kl}.
\label{full_S}
\end{eqnarray}
Once $\Pi_{kl}$ and $\Sigma_{kl}$ are evaluated at the $n$-loop level, 
$D_{kl}$ and $S_{kl}$ are obtained at the same order.

The effective masses  $m^B_{\rm eff}$ and $m^F_{\rm eff}$ are read from the large distance behavior of 
$D_{kl}$ and  $S_{kl}$: For $|k-l|\gg 1$,
\begin{eqnarray}
&&D_{kl} \approx C e^{-m^B_{\rm eff}|k-l|},  \\
&& S_{kl} \approx C^\prime \theta_{k,l} e^{-m^F_{\rm eff}|k-l|}.
\end{eqnarray}
with
\begin{eqnarray}
\theta_{k,l} \equiv
\left\{
\begin{array}{ll}
 1 &\quad {~~\rm for}~ k\ge l \\
 0 &\quad {~~\rm for}~ k<l. 
\end{array}
\right.
\label{theta}
\end{eqnarray}
At one-loop level, the self-energies provide the shifts of masses $\Delta m$
in $D_0^{-1}(p)$ and $S_0^{-1}(p)$ via (\ref{full_D}) and (\ref{full_S}).
The one-loop effective masses $m^{B,F}_{\rm eff}$ 
are actually obtained from the formulas of tree level effective masses 
$m^{B,F}_{0,\rm eff}$ with $m\rightarrow m+ \Delta m$.

\subsection{The naive action}

We begin with the case of the naive action (\ref{naive_action}) 
whose $D_0 (p)$ and $S_0 (p)$  are given by
\begin{eqnarray}
&&  D_0 (p) \equiv \frac{1}{2(1-\cos p) + m^2},\\
&&  S_0 (p) \equiv \frac{1}{1-e^{-ip} + m}.
\end{eqnarray}  
The free boson propagator in the position space is evaluated from (\ref{boson_propagator}): 
\begin{eqnarray}
D_{0, kl} = \oint dz \frac{z^{k-l}}{z^2 -(m^2+2) z+1},
\label{boson_propagator_naive_int}
\end{eqnarray}
for  $z=e^{ip}$.
It is easily shown that 
\begin{eqnarray}
D_{0, kl} = \frac{e^{- m_{0, \rm eff}^B |k-l|}}{2m\sqrt{1+\frac{m^2}{4}}},
\label{boson_propagator_naive}
\end{eqnarray}
where 
\begin{eqnarray}
m^B_{0, \rm eff} = -{\rm log} \left(1+\frac{m^2}{2}-m\sqrt{1+\frac{m^2}{4}} \right).
\label{mass_boson_naive_leading}
\end{eqnarray}
Similarly,  
\begin{eqnarray}
S_{0, kl} = \theta_{k,l}
\frac{e^{-m_{0, \rm eff}^F|k-l|}}{1+m}
\label{fermion_propagator_naive}
\end{eqnarray}
where 
\begin{eqnarray}
m^F_{0, \rm eff} = {\rm log} \left( 1+m \right),
\label{mass_fermi_naive_leading}
\end{eqnarray}
and $\theta_{k,l}$ is given by (\ref{theta}).

At one-loop level,  the boson and fermion self energies are obtained as
\begin{eqnarray}
&&\Pi (p)=6\lambda m^2(\frac{1}{\sqrt{1+\frac{m^2}{4}}}-\frac{1}{1+m}), 
\label{naive_Pi}\\
&& \Sigma (p)=\frac{3\lambda m}{2\sqrt{1+\frac{m^2}{4}}}.
\label{naive_Sigma}
\end{eqnarray}
The one-loop self energies provide different corrections to the mass 
$m \rightarrow m + \Delta m_{B,F}$ where $\Delta m_B$ and  $\Delta m_F$ are
 identified from (\ref{naive_Pi})
and  (\ref{naive_Sigma}), respectively. 

The one-loop effective masses are obtained by inserting $m+\Delta m_{B,F}$ 
into (\ref{mass_boson_naive_leading}) and (\ref{mass_fermi_naive_leading}):
\begin{eqnarray}
&&\frac{E_1^{B}}{m}=1+
3\lambda ma + \frac{(-1-81\lambda)m^2a^2}{24} + O(\lambda^2,m^3a^3)\\
&&\frac{E_1^{F}}{m}=1+\frac{3\lambda}{2} 
-\frac{(2+6\lambda)ma}{4} + O(\lambda^2,m^2a^2) .
\end{eqnarray}
We should note that $E_1^{B}$ is different from  $E_1^{F}$ even in the continuum limit $ma \rightarrow 0$
as a result of the one-loop effect although they coincide with each other 
at the tree level with $\lambda=0$.

\subsection{The CG action}

The free propagators of the CG action are 
\begin{eqnarray}
&& D^{CG}_0(p) = \frac{1}{2(1-\cos p) + m^2+2m(1-\cos p)},\\
&& S^{CG}_0(p) = \frac{1}{1-e^{-ip} + m}.
\label{CG-propagator}
\end{eqnarray} 
The similar calculation as done around (\ref{boson_propagator_naive_int}) tells us that 
the effective masses are degenerated as   
\begin{equation}
m_{0,\rm eff}^B=
m_{0,\rm eff}^F= {\rm log}(1+m),
\label{emass_tree_CG}
\end{equation}
at the tree level.

The self energies are calculated at the one-loop level as 
\begin{eqnarray}
&&\Pi(p) = \Delta m [2m + 2(1-{\rm cos}p)], \\
&&\Sigma(p) = \Delta m, 
\end{eqnarray}
where 
\begin{eqnarray}
\Delta m \equiv \frac{3\lambda m}{2+m}.
\end{eqnarray}
These give the same correction to the boson mass and the fermion mass up to $O(\lambda)$.
The one-loop effective masses are evaluated from (\ref{emass_tree_CG}) with $m + \Delta m$.
We  thus obtain that
\begin{equation}
\frac{E_1}{m}=1+\frac{3\lambda}{2} -\frac{ma}{2}-\frac{\lambda (ma)^2}{2}
-\frac{9\lambda (ma)^2}{4} + O(\lambda^2,(ma)^3),
\end{equation}
for $E_1 \equiv m^B_{\rm eff}=m^F_{\rm eff}$ owing to an exact SUSY.

\subsection{The CLR action}

The free propagators of the CLR action are given by
\begin{eqnarray}
& \displaystyle{D_{0}^{CLR}(p)\equiv\frac{1}{2(1-\cos p) + m^2(1+\cos p)/2},}\\
& \displaystyle{S_{0}^{CLR}(p)\equiv\frac{1}{1-e^{-ip} + m(1+e^{-ip})/2}} .
\end{eqnarray} 
The tree level effective masses are
\begin{equation}
m_{\rm eff}^B=
m_{\rm eff}^F= {\rm log}\left(\frac{1+\frac{m}{2}}{1-\frac{m}{2}}\right),
\label{emass_tree_CLR}
\end{equation}
which are degenerated between the boson and the fermion.

The one-loop self energies are given by 
\begin{eqnarray}
&& \Pi(p)=2m \Delta m (1+{\rm cos} p), 
\\
&&\Sigma (p)= \Delta m \left(\frac{1+e^{-ip}}{2}\right),
\end{eqnarray}
where 
\begin{eqnarray}
\Delta m= \frac{\lambda m (m+6)}{2(m+2)}.
\end{eqnarray}
The one-loop effective masses are read from (\ref{emass_tree_CLR}) with $m+\Delta m$.
The first excited energies for the bosonic and fermionic states  are thus obtained 
as $E_1 \equiv m^{B}_{\rm eff}=m^{F}_{\rm eff}$:
\begin{equation}
\frac{E_1}{m} = 1+ \frac{3\lambda}{2} -\frac{\lambda ma}{2} +\frac{(ma)^2}{12}
+\frac{5\lambda (ma)^2}{8}+O(\lambda^2,(ma)^3),
\end{equation}
owing to an exact SUSY.